\documentclass{aastex631}
\submitjournal{ApJ}

\usepackage{booktabs}  
\usepackage{array}     
\usepackage{graphicx}
\usepackage{tabularx}
\usepackage{amsmath}
\usepackage{amssymb}
\usepackage{booktabs}       
\usepackage{hyperref}
\usepackage{savesym}
\savesymbol{tablenum}
\usepackage{siunitx}
\usepackage{xcolor}
\usepackage{orcidlink}
\usepackage{makecell}
\usepackage{subfigure}

\restoresymbol{SIX}{tablenum}

\begin{document}

\title{
	Disentangling Stellar Mass and Environmental Effects on eROSITA AGN Activity using multi-wavelength data from GAMA, WISE, GALEX, and DESI Legacy Survey
}

\author{Navya Saraswat \orcidlink{0009-0002-1998-9042}}
\affiliation{Department of Physics, IIT Hyderabad, Kandi, Telangana 502284, India; \href{mailto:ph26resch01008@iith.ac.in}{ph26resch01008@iith.ac.in}}

\author{Shadab Alam \orcidlink{0000-0002-3757-6359}}
\affiliation{Tata Institute of Fundamental Research, Homi Bhabha Road, Mumbai 400005, India}

\author{Subhabrata Majumdar \orcidlink{0009-0009-4972-6738}}
\affiliation{Tata Institute of Fundamental Research, Homi Bhabha Road, Mumbai 400005, India}

\date{\today}

\begin{abstract}
	We use a stellar-mass complete sample of $\sim\!36,000$ galaxies from GAMA/eFEDS to test whether AGN activity depends on environment independent of stellar mass. To isolate this effect, we introduce $\delta_{\rm rank}$, a novel local over-density metric defined within narrow stellar mass bins. Our main result is that X-ray--selected AGNs show a clear environmental trend, the AGN fraction rises from $\sim\!4\%$ in low-density regions to $\sim\!6\%$ in high-density regions, with a sharp transition at $\delta_{\rm rank}\sim\!0.4$.
	This environmental dependence is absent for WISE-selected and broad-line AGN, and Eddington ratios show no variation across environments. We also found that for galaxies with SFR $> 15\,M_\odot\,\mathrm{yr}^{-1}$, X-ray AGN fraction is  approximately 5\%, 60\% and 20\% in low, mid and high $\delta_{\rm rank}$ galaxies. 
	We show that relationship between AGN fraction and SFR varies dramatically with both environment and selection method. 
	We conclude that environment on scales of $\sim\!5\mathrm{h}^{-1}$~Mpc regulates X-ray AGN triggering efficiency.
\end{abstract}


\section{Introduction}
\label{sec:intro}

The luminous manifestation of accretion onto supermassive black
holes (SMBHs) at the centres of galaxies is known as active galactic nuclei (AGN). AGN play a pivotal role in regulating star formation and
driving galaxy evolution across cosmic time \citep{Kormendy2013, Heckman2014}. The tight
correlations observed between SMBH mass and host-galaxy properties, such as bulge mass, stellar
velocity dispersion, and total stellar mass suggest a co-evolutionary relationship between black
holes and their host galaxies \citep{Magorrian1998, Gebhardt2000, Ferrarese2000, Tremaine2002}.
Understanding what triggers and sustains AGN activity is therefore fundamental to building a
complete picture of galaxy evolution.

The fraction of galaxies hosting an actively accreting AGN has been shown to depend strongly on
host-galaxy stellar mass \citep{Kauffmann2003, Best2005, Aird2012}. More massive galaxies are
significantly more likely to harbour X-ray luminous AGN, a trend that holds across a wide range
of redshifts and AGN selection techniques. \citet{Aird2012} demonstrated that the probability of
a galaxy hosting an AGN above a given specific accretion rate follows a distribution that is
largely independent of star formation rate (SFR) once stellar mass is accounted for, suggesting that
the primary driver of AGN activity is the availability of a sufficiently massive black hole rather
than the current mode of star formation in the host. This stellar mass dependence, if not
carefully controlled for, can obscure genuine environmental signals,
making it essential to disentangle these two effects before drawing conclusions about environmental
influences on nuclear activity.

Galaxy environment shapes the observable properties of galaxies at all epochs. Observational
studies have established that denser environments tend to host older, redder, and more quiescent
stellar populations \citep{Dressler1980, Peng2010}, a trend attributed to a combination of
internal processes (AGN and stellar feedback) and external mechanisms such as ram-pressure
stripping, tidal interactions, galaxy harassment, and strangulation of the gas supply
\citep{Boselli2006}. The same environmental processes that regulate star formation may also
modulate the supply of cold gas to galaxy nuclei and hence influence the frequency and character
of AGN activity.

The relationship between AGN activity and large-scale environment has been the subject of
extensive investigation, with conflicting results in the literature. Some studies find that AGN
preferentially avoid the densest cluster environments \citep{Kauffmann2004},
consistent with the picture that the hot intra-cluster medium removes cold gas from in-falling
galaxies before it can fuel the central black hole. Others report enhanced AGN incidences in
intermediate-density environments such as galaxy groups and cluster outskirts, where tidal
interactions and galaxy-galaxy mergers may drive gas inflows toward nuclear regions
\citep{Silverman2009, Arnold2009}. The lack of consensus in
the literature likely reflects a combination of differing AGN selection techniques, sample sizes,
redshift ranges, and, critically, inadequate control for the confounding influence of stellar
mass.

Different AGN selection methods are sensitive to distinct physical processes and AGN populations,
and this selection dependence may itself carry environmental information. X-ray surveys probe
both obscured and unobscured AGN through their high-energy emission, largely independent of
dust \citep{Brandt2015}. Mid-infrared (IR) colour selection using data from the Wide-field
	Infrared Survey Explorer \citep[WISE;][]{Wright2010} traces warm dust heated by the
AGN torus and is effective at identifying luminous, dust-obscured systems \citep{Stern2012,
	Assef2013, Hviding2022}. Optical spectroscopic diagnostics, including broad emission lines
(Type~I AGN) and narrow-line BPT diagnostics \citep{Baldwin1981} are sensitive to the
ionisation state of the narrow-line region and the presence of an unobscured broad-line region (BLR).
Since these selection techniques trace AGN at different luminosities, obscuration levels, and
accretion rates, the environmental dependencies they reveal need not agree, and comparing their
behavior provides a powerful lever for identifying the physical origin of any environmental
signal.

The construction of large, spectroscopically complete galaxy surveys at low redshift, combined
with deep multi-wavelength imaging and sensitive X-ray observations, has created new
opportunities to address these questions with statistical rigor. The Galaxy and Mass Assembly
\citep[GAMA;][]{Driver2011, Driver2022} survey provides the spectroscopic foundation of this work 
and the three-dimensional density fields from \citet{Alam2018} are used to characterise the local environment of galaxies.
The eROSITA Final Equatorial-Depth Survey \citep[eFEDS;][]{Brunner2022} provides sensitive X-ray coverage over the GAMA equatorial fields, probing nuclear activity independent of line-of-sight obscuration. Combining these datasets with ultraviolet (UV) photometry from the Galaxy Evolution Explorer \citep[GALEX;][]{Martin2005} survey, deep optical imaging from the DESI Legacy Imaging Survey \citep{Dey2019},
mid-IR photometry from WISE, and high-resolution optical
spectroscopy from DESI DR1 \citep{DESICollaboration2024, DESI2025} creates an exceptionally rich
multi-wavelength baseline for comprehensive spectral energy distribution (SED) modeling and
AGN identification.

A central methodological challenge in environment--AGN studies is isolating the environmental
effect from the stellar mass dependence of the AGN fraction. Since massive galaxies preferentially
reside in denser environments \citep{Peng2010, Darvish2016}, any observed increase in the AGN
fraction with density could simply reflect the underlying mass--environment relation rather than
a genuine environmental influence on nuclear activity. Previous studies have addressed this by
simultaneously binning in stellar mass and environment, but this approach is limited by small
number statistics and does not fully remove residual correlations. In this paper we introduce a  ranking parameter, $\delta_\mathrm{rank}$, that measures the relative overdensity of each
galaxy compared to other galaxies of similar stellar mass, effectively removing the
mass--environment correlation by construction and enabling a clean measurement of any genuine
environmental dependence.

In this work, we exploit the GAMA--eFEDS overlap to study the environmental dependence of AGN
activity in a stellar mass-complete sample of $\sim\!$ 36,000 galaxies at $z \leq 0.4$. We
perform comprehensive SED fitting using \textsc{cigale} \citep{Boquien2019} to derive stellar
masses and SFRs, identify AGN through four independent and complementary
techniques, quantify the three-dimensional environment around each galaxy using the Voronoi
tessellation and tidal tensor formalism of \citet{2018MNRAS.476.5442P} and \citet{Alam2018}, and examine how the AGN fraction varies
with environment after rigorously controlling for stellar mass. We further investigate the
relationship between SFR, environment, and AGN activity, and test whether
environment modulates only the triggering efficiency of AGN or also the subsequent accretion
physics as quantified by the Eddington ratio distribution.

This paper is organized as follows. Section~\ref{sec:data} describes the multi-wavelength
datasets, the cross-matching procedure, the SED fitting methodology, and the construction of the
stellar mass-complete sample. Section~\ref{sec:agn_identification} presents the four AGN identification
methods and characterises each sample. Section~\ref{sec:environment} describes the environmental
measurements, the $\delta_\mathrm{rank}$ methodology, and robustness tests against scale
dependence. Section~\ref{sec:results} presents our main results on the stellar mass and
environmental dependence of the AGN fraction, the SFR--environment--AGN connection, and the
Eddington ratio distributions. Section~\ref{sec:conclusions} summarizes our principal conclusions, and in Section~\ref{sec:data_avail}, we describe the public availability of the analysis pipeline and value-added data products associated with this work.

Throughout this work, we assume a flat $\Lambda$CDM cosmology with $H_0 = 70$\,km\,s$^{-1}$\,Mpc$^{-1}$,
$\Omega_\mathrm{m} = 0.3$, and $\Omega_\Lambda = 0.7$. Stellar masses and SFRs
are quoted assuming a \citet{Chabrier2003} initial mass function. 

\section{Data and Sample Selection}
\label{sec:data}

Our analysis draws on six independent surveys spanning from the UV to the X-ray, unified by their overlap with the G09 equatorial field of the GAMA survey. We describe the assembly of this multi-wavelength dataset below, followed by the cross-matching strategy, SED fitting procedure, and the criteria used to define a stellar mass-complete working sample.

\subsection{Parent Galaxy Sample and Multi-wavelength Data}
\label{sec:parent_sample}

We construct our parent galaxy sample using the GAMA \citep{Driver2011, Driver2022}  survey, a highly complete low-redshift spectroscopic survey covering 
the G09, G12, and G15 equatorial fields. These regions benefit 
from extensive multi-wavelength coverage and provide reliable redshifts and optical photometry 
for nearly 300,000 galaxies. For this study, we restrict the sample to the overlapping footprint 
of GAMA, eFEDS, the DESI Legacy Imaging Survey, and GALEX, and to sources with redshift 
$z \leq 0.4$. The final GAMA G09 catalog contains 82,837 objects over an area of approximately $\sim\!60~\mathrm{deg}^2$. The six surveys contributing to 
this analysis and their primary roles are summarised in Table~\ref{tab:surveys}.

\begin{table*}
	\centering
	
	\begin{tabular*}{\textwidth}{@{\extracolsep{\fill}} l l l}
		\toprule
		\textbf{Survey} & \textbf{Bands} & \textbf{Primary Role} \\
		\midrule
		
		GAMA &
		Optical spectroscopy &
		Spectroscopic redshifts, emission-line measurements \\
		
		DESI Legacy &
		$g$, $r$, $z$ &
		Deep optical photometry, stellar mass estimation \\
		
		GALEX &
		FUV, NUV &
		UV flux densities, star formation tracer \\
		
		WISE &
		$W1$, $W2$ &
		Mid-IR photometry, AGN-heated dust \\
		
		eFEDS &
		$0.2$--$2.3$ keV &
		Soft X-ray coverage, AGN detection \\
		
		DESI DR1 &
		Optical spectroscopy &
		Emission-line fitting, black hole mass estimation
		\citep{Shrivastav2026} \\
		
		\bottomrule
	\end{tabular*}
	
	\caption{Summary of surveys used in this analysis.}
	\label{tab:surveys}
	
\end{table*} 

\begin{figure*}
	\centering
	
	\includegraphics[width=0.9\textwidth]{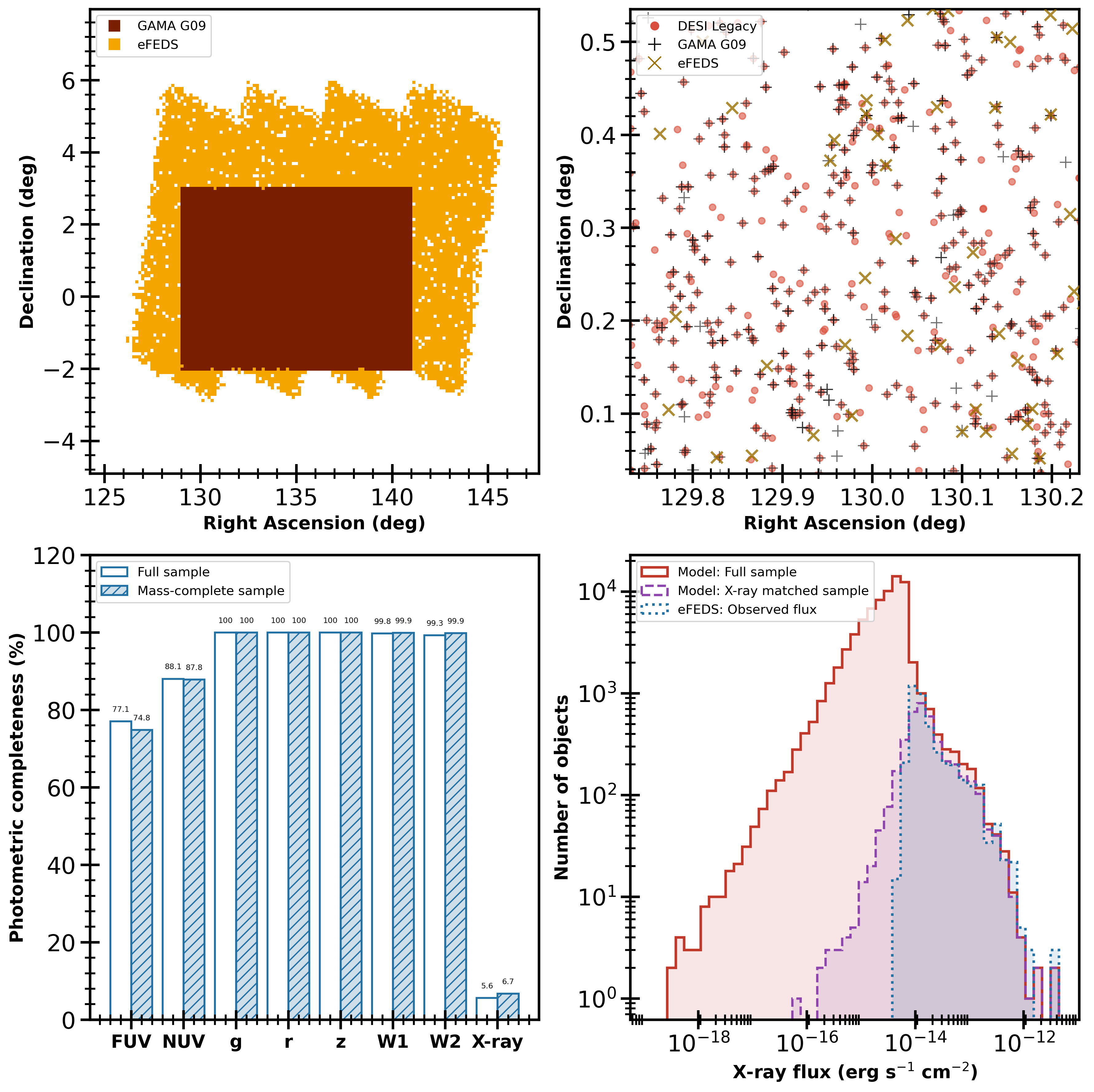}
	\caption{Survey coverage, sample characterisation, and photometric completeness for the GAMA–eFEDS   matched sample. \textbf{Top left:} Sky footprints of the GAMA G09 (dark brown) and eFEDS (orange) surveys, highlighting their overlapping coverage. \textbf{Top right:} A $0.5 \times 0.5$ deg$^2$ region centred on the most massive galaxy in the sample, showing DESI (red circles), eFEDS (gold crosses), and GAMA (black plus symbols) sources. DESI provides optical counterparts for most GAMA galaxies, while a population of X-ray sources lacking optical identifications is also apparent.
		\textbf{Bottom left:} Photometric completeness across wavebands for the full sample (open bars) and the stellar mass-complete subsample (filled bars), spanning FUV, NUV, optical ($g$, $r$, $z$), mid-IR ($W1$, $W2$), and X-ray bands. Values indicate the fraction of sources with reliable photometric measurements in each band.
		\textbf{Bottom right:} Comparison of X-ray flux distributions from CIGALE (Code Investigating GALaxy
		Emission) predictions and eFEDS observations. The red histogram shows model-predicted fluxes for the full optical sample, the purple histogram shows predictions for eFEDS-matched galaxies, and the blue histogram shows the observed eFEDS fluxes. The sharp decline at $\sim10^{-14},\mathrm{erg,s^{-1},cm^{-2}}$ reflects the eFEDS detection threshold, illustrating the flux limit that governs the X-ray-selected sample.
	}
	\label{fig:multipanel-1}
\end{figure*}

Figure~\ref{fig:multipanel-1}, in the top left panel, shows the sky overlap between these surveys focused around the GAMA G09 region. The top-right panel shows a zoomed-in view of the most massive galaxy highlighting the structure seen by various surveys. 
Each of these surveys provides complementary information about galaxies, and combining them through cross-matching enables a more complete physical understanding of the galaxies we detect. 
This is particularly important because each survey is carried out independently and often with different resolutions, sensitivities, 
and field-of-view geometries. An X-ray detection indicates a high-energy source, but without an 
optical/IR counterpart or redshift its nature is ambiguous. By matching to deep optical/IR 
imaging and spectroscopy, one can classify sources as AGN, normal galaxies, or stars, and 
determine distances. Combining X-ray flux with optical spectral lines and photometry therefore 
allows the calculation of AGN accretion rates, host-galaxy masses, and SFRs.

\subsection{Catalog Cross-matching}
\label{sec:crossmatch}

We associate detections of the same galaxy across different surveys via positional cross-matching on the celestial sphere. While more sophisticated probabilistic cross-identification frameworks exist \citet{Budavari2008}, positional matching remains the standard approach for large survey analyses and is well-suited to our dataset given the survey depths and astrometric precisions involved.

For each object in a given catalog, we compute the angular separation ($\theta$) from all sources in the comparison catalog using the small-angle approximation

\begin{equation}
	\theta \approx \sqrt{ \left[ (\alpha_1 - \alpha_2)\cos\delta_1 \right]^2 + (\delta_1 - \delta_2)^2 } \, ,
\end{equation}

\noindent where $(\alpha_1, \delta_1)$ and $(\alpha_2, \delta_2)$ are the right ascension and declination of the two sources, respectively. The nearest neighbor is retained as a potential counterpart if the separation lies within a predefined tolerance of $10^{-3}$ radians (corresponding to $\sim\!3.4^\prime$). The survey-specific criteria used to identify reliable counterparts are described in the following subsections.

\subsubsection{GAMA$\times$eFEDS}
\label{sec:xray_crossmatch}

We cross-match X-ray sources from eFEDS (described in Section~\ref{sec:xrayflux}) with galaxies in the GAMA G09 field. An initial sky-based positional match yields $74{,}493$ candidate associations from $27{,}910$ eFEDS sources. On its own, this number is not very meaningful. The positional uncertainties in eFEDS are relatively large (median $\approx 4.7$ arcsec), so using a fixed angular cutoff would treat well-constrained and poorly constrained sources in the same way.

We instead normalise each angular separation by the corresponding positional uncertainty, $\sigma_\theta$, reported in the eFEDS catalog, and retain only those pairs satisfying $d_\sigma = \theta / \sigma_\theta \leq 5$. This threshold is motivated by the observed distribution of $d_\sigma$, which rises steeply at small values, where true counterparts are expected, and flattens beyond $d_\sigma \approx 5$,  indicating a population dominated by chance alignments. Applying this cut reduces the sample to 4{,}892 X-ray–optical associations. This corresponds to $\sim\!5.9\%$ of GAMA sources having X-ray counterparts within the adopted threshold. The resulting sample constitutes our X-ray-detected galaxy population.

\subsubsection{GAMA$\times$DESI Legacy}

The DESI Legacy Imaging Surveys \citep{Dey2019} are a wide-area optical imaging program designed to support target selection for the Dark Energy Spectroscopic Instrument (DESI). The survey combines observations from the Dark Energy Camera Legacy Survey (DECaLS,\cite{DECAM2015}), the Beijing-Arizona Sky Survey (BASS, \cite{BASS2017}), and the Mayall z-band Legacy Survey (MzLS, \cite{MZLS2016}), providing deep imaging in the $g$, $r$, and $z$ bands over a large fraction of the extragalactic sky. The Legacy Surveys reach significantly greater depth than previous wide-field optical surveys and provide uniform photometric measurements across multiple bands.

We restrict the Legacy catalog to the G09 footprint and cross-match it with the GAMA spectroscopic catalog using a positional tolerance of 1 arcsecond, appropriate given the sub-arcsecond astrometric accuracy of both surveys. This procedure recovers almost all sources, confirming the near-complete spatial overlap between the two datasets. For subsequent analysis, we adopt the Legacy forced photometry for all matched galaxies.

\subsection{Multi-wavelength Photometry}
\label{sec:photometry}

Having established the matched galaxy sample, we next compile the photometric measurements required for physical characterisation of the sources. Since galaxies emit across a broad range of wavelengths, observations at different wavelengths capture different physical components; UV emission traces recent star formation, optical and near-IR light mainly originates from stellar populations, IR emission is associated with dust reprocessing, and X-rays probe high-energy processes such as accretion onto SMBHs and shocked hot gas.

The resulting wavelength coverage and per-band completeness of our assembled 
photometry are summarised in the bottom-left panel of Figure~\ref{fig:multipanel-1}. 
All flux measurements are converted to flux density units of mJy prior to analysis. 
This broad spectral coverage enables robust SED fitting, from which the relative 
contributions of stellar, dust, and AGN components can be systematically 
decomposed, as described in Section~\ref{sec:cigale}.

\subsubsection{Optical and IR Photometry}

Optical ($g$, $r$, $z$) and mid-IR ($W1$, $W2$) fluxes are drawn from the DESI Legacy Imaging Survey 
and WISE respectively. WISE is a NASA all-sky survey mission that mapped the entire sky at 3.4, 4.6, 12, and 22~$\mu$m ($W1$--$W4$) with angular 
resolutions of 6--12 arcseconds, reaching 5$\sigma$ point-source depths of approximately 0.08 and 
0.11~mJy in the $W1$ and $W2$ bands; we use forced photometry from the Legacy Survey source catalog.

These five bands form 
the backbone of the SED analysis; the optical traces the light-weighted stellar population, while the near-IR, being relatively insensitive to dust attenuation, provides a robust tracer for stellar mass. The $W1$ band in particular is well suited to this role, given its low dust opacity and minimal contamination from AGN-heated dust, which becomes significant only at $W2$ and longer wavelengths for moderately luminous AGN.

Both catalogs report fluxes in nanomaggies, converted to mJy via 
$F_\nu = F_{\mathrm{nmgy}} \times 3.631 \times 10^{-3}$, with uncertainties 
derived from the inverse square root of the catalog inverse-variance values. 
Legacy photometry is measured using the \texttt{Tractor} source extraction code 
\citep{Lang2016}, which fits point spread function (PSF) and galaxy profiles 
simultaneously, ensuring reliable flux measurements for both point sources and 
extended low-surface-brightness emission.

\subsubsection{UV Photometry}

GALEX \citep{GALEX2005} is a NASA space mission that conducted imaging surveys of the sky in two 
bands from 2003 to 2013: the Near-UV (NUV; 1750--2750\,\AA) and Far-UV 
(FUV; 1350--1750\,\AA) channels. Coverage of the GAMA equatorial fields is 
provided through the Medium Imaging Survey, reaching limiting AB magnitudes of 
approximately 23.0 and 22.6 in the NUV and FUV respectively, at an angular 
resolution of ${\sim}5$ arcseconds.

These two bands extend the SED to wavelengths dominated by emission from massive 
stars with lifetimes shorter than ${\sim}100$ Myr, making them direct tracers of 
recent star formation. The two channels are not redundant: the FUV is sensitive to 
star formation on timescales of $\lesssim 10$ Myr, while the NUV traces longer 
timescales, and their ratio encodes information on recent star formation history 
that is inaccessible from optical data alone.

In AGN host galaxies, these bands serve an additional diagnostic purpose. 
Accretion onto the central black hole can produce an excess at short wavelengths 
that mimics or exceeds the stellar contribution; without such constraints, this 
degeneracy can bias SFR estimates derived from SED fitting. Where detected, this 
excess emission can be attributed to the AGN component rather than to stellar 
processes. NUV coverage is available for 88\% of the sample and FUV for 77\%, 
with the incompleteness reflecting the shallower depth and less uniform sky 
coverage relative to the optical surveys. Sources without detections are included 
using upper limits where available.

\subsubsection{X-ray Flux Conversion}
\label{sec:xrayflux}

X-ray fluxes are taken from eFEDS, a survey conducted during the performance verification phase of Spektrum-Roentgen-Gamma \citep[SRG/eROSITA;][]{Predehl2021}. Covering approximately $140\,\mathrm{deg}^2$ across the GAMA equatorial fields with a typical exposure time of $\sim$ 2.2\,ks, eFEDS represents one of the deepest wide-area X-ray surveys available at the time of these observations, reaching a point-source sensitivity $\sim\!10^{-14}$\,erg\,s$^{-1}$\,cm$^{-2}$ in the soft band. Source detection was carried out using a sliding-cell maximum-likelihood algorithm, with fluxes 
measured in the 0.2--2.3\,keV band and corrected for vignetting, point spread function losses, and Galactic absorption. For AGN selection in  Section~\ref{sec:xray_agn}, we use maximum-likelihood fluxes measured in the 0.5--2\,keV band, which corresponds to the completeness-characterised interval 
from \citet{Brunner2022}.

The soft-band selection influences the AGN population that is recovered. Heavily obscured AGN, for which the intrinsic X-ray emission is absorbed below 
$\sim$ 2\,keV, are systematically underrepresented, biasing the X-ray-detected subsample toward unobscured or moderately obscured systems. This limitation 
should be considered when interpreting AGN fractions based solely on X-ray detections, and motivates the use of optical and IR diagnostics as 
complementary AGN indicators for the X-ray-undetected majority.

The flux-to-energy conversion adopted here assumes a flat spectrum response across the 0.2--2.3\,keV band (a boxcar approximation), reasonable for a power-law AGN 
spectrum with photon index $\Gamma \sim 1.7$--$2.0$. For sources with significantly harder or softer spectra this introduces uncertainties at the 
$\sim$ 10--20\% level, which is small relative to the constraints from optical and IR data spanning several orders of magnitude in frequency and does 
not significantly affect the derived physical parameters.

\subsection{SED Fitting with CIGALE}
\label{sec:cigale}

We fit the assembled multi-wavelength photometry using the Code Investigating GALaxy Emission \citep[CIGALE;][]{Boquien2019}, which is an energy-balance SED fitting framework that models the emission from galaxies across the UV, optical, IR, and X-ray bands. By combining physically motivated models for stellar populations, dust attenuation and re-emission, nebular emission, and AGN activity, CIGALE enables the simultaneous estimation of galaxy and AGN properties from broadband photometric data. This approach is especially important for AGN host galaxies, where nuclear emission can significantly affect measurements of stellar mass and SFR if not explicitly accounted for.

\subsubsection{CIGALE Configuration}
\label{sec:cigale_config}

CIGALE constructs a grid of model SEDs by combining physical modules describing the star formation history, stellar populations, dust attenuation, dust emission, AGN torus emission, and X-ray emission. For each source, the observed photometry is compared against the model grid, and physical parameters are derived through Bayesian marginalisation over all acceptable models. Posterior estimates of stellar mass, SFR, dust attenuation, and AGN contribution are obtained together with their associated uncertainties.

The module configuration adopted here includes a delayed exponentially declining 
star formation history (\texttt{sfhdelayed}), stellar population synthesis models 
from \citet{Bruzual2003} (\texttt{bc03}), nebular emission from H\,\textsc{ii} 
regions (\texttt{nebular}), a modified starburst dust attenuation law 
(\texttt{dustatt\_modified\_starburst}) coupled with the IR emission 
templates of \citet{Dale2014} (\texttt{dale2014}), the SKIRTOR AGN torus models 
of \citet{Stalevski2016} (\texttt{skirtor2016}), and the X-ray module of 
\citet{Yang2020} (\texttt{yang20}).

This combination is intended to capture the full range of physical processes contributing to the SED, from recent star formation to AGN-heated dust and X-ray emission. Stellar population parameters were fixed as follows: a \citet{Chabrier2003} IMF, solar metallicity ($Z = 0.02$), and nebular ionisation parameter $\log U = -2.0$. For the AGN torus module, we fix the SKIRTOR optical depth $\tau_{9.7} = 7$, 
dust grain power-law index $p = 1.0$, polar index $q = 1.0$, half-opening angle $40^\circ$, outer-to-inner radius ratio $R = 20$, and clump mass fraction $M_\mathrm{cl} = 0.97$. X-ray emission parameters are fixed at photon index $\Gamma = 1.8$ and high-energy cutoff $E_\mathrm{cut} = 300\,\mathrm{keV}$. The free parameters explored during the fitting are summarised in 
Table~\ref{tab:cigale_module}.

 Physical parameters are estimated using the \texttt{pdf\_analysis} method, which integrates over the full model grid to obtain likelihood-weighted posterior distributions.

\begin{table*}
	\centering

	\begin{tabular*}{\textwidth}{@{\extracolsep{\fill}} l l l c}
		\toprule
		\textbf{Parameter} & \textbf{Symbol} & \textbf{Values explored} & \textbf{Count} \\
		\midrule
		
		\multicolumn{4}{c}{\textbf{Star formation history}} \\
		e-folding time &
		$\tau_{\rm main}$ &
		100, 500, 1000, 5000, 10000 Myr &
		5 \\
		
		Main population age &
		$t_{\rm main}$ &
		500, 1000, 3000, 5000, 7000, 10000 Myr &
		6 \\
		
		\midrule
		
		\multicolumn{4}{c}{\textbf{Dust attenuation}} \\
		Nebular colour excess &
		$E(B-V)_{\rm lines}$ &
		0.00, 0.05, 0.10, 0.20, 0.30, 0.50 mag &
		6 \\
		
		Attenuation slope &
		$\delta$ &
		$-1.3$, $-0.7$, $-0.4$, $0.0$ &
		4 \\
		
		\midrule
		
		\multicolumn{4}{c}{\textbf{IR dust emission}} \\
		Heating slope &
		$\alpha$ &
		1.5, 2.0, 2.5 &
		3 \\
		
		\midrule
		
		\multicolumn{4}{c}{\textbf{AGN torus}} \\
		Viewing angle &
		$i$ &
		$30^\circ$ (Type~I), $70^\circ$ (Type~II) &
		2 \\
		
		AGN IR fraction &
		$f_{\rm AGN}$ & 0.01, 0.10–0.90 (steps of 0.04), 0.99 &
		23 \\
		
		\midrule
		
		\multicolumn{4}{c}{\textbf{X-ray emission}} \\
		UV-to-X-ray slope &
		$\alpha_{\rm ox}$ &
		$-1.9$ to $-1.1$ in steps of 0.1 &
		9 \\
		
		Inclination coefficient &
		$a_{\rm inc}$ &
		0.0, 0.5 &
		2 \\
		
		\midrule
		
		\multicolumn{4}{c}{Total grid size:
			$5\times6\times6\times4\times3\times2\times23\times9\times2
			\approx 1.8\times10^{6}$ models} \\
		
		\bottomrule
	\end{tabular*}
	\caption{Parameter grid used in the spectral energy distribution (SED)
		fitting with CIGALE. The table lists,
		for each module, the parameter name and the values or range sampled during
		the fit. Fixed parameters are not listed here; see
		Section~\ref{sec:cigale_config} for their values.}
	\label{tab:cigale_module}
	
\end{table*}

\subsubsection{Filter Setup and X-ray Integration}

Filter transmission curves for the WISE $W1$ and $W2$ and GALEX FUV and NUV 
bands are available natively in the CIGALE database. Curves for the Legacy 
$g$, $r$, and $z$ bands are taken from the Data Release 8 photometry 
description of the DESI Legacy Imaging Surveys\footnote{\url{https://www.legacysurvey.org/dr8/description/}} and added 
manually. For the X-ray regime, we define a custom boxcar filter spanning 
the eFEDS energy range of 0.2--2.3\,keV, in line with the flat spectral 
assumption adopted during flux conversion (Section~\ref{sec:xrayflux}). 

\subsubsection{X-ray Upper Limits}

The majority of sources in the optical sample are not detected by eFEDS because their fluxes fall below the survey sensitivity limit. Excluding these sources from the X-ray analysis would discard useful information, while treating them as true zero-flux measurements would bias the inferred AGN properties. Instead, we assign upper limits and incorporate them into CIGALE via negative flux uncertainties, allowing non-detections to provide one-sided constraints without contributing positive flux measurements to the likelihood calculation. The upper limit is set at $F_{X,\mathrm{lim}} = 7.6 \times 10^{-15}\,\mathrm{erg\,s^{-1}\,cm^{-2}}$, 
derived from the observed flux distributions shown in the bottom-right panel of Figure~\ref{fig:multipanel-1}. Both the eFEDS detections and the corresponding CIGALE model predictions for matched sources peak in the range $\sim\!10^{-14}$--$10^{-13}\ \mathrm{erg\,s^{-1}\,cm^{-2}}$ and show a sharp truncation just above $\sim\!10^{-14}\ \mathrm{erg\,s^{-1}\,cm^{-2}}$, which reflects the effective sensitivity limit of the survey. In contrast, model predictions for the full optical sample extend to significantly lower fluxes, indicating that the undetected population lies below this threshold rather than being intrinsically X-ray faint.
\subsubsection{Fit Quality}

\begin{figure*}
	\centering
	\includegraphics[width=0.9\textwidth]{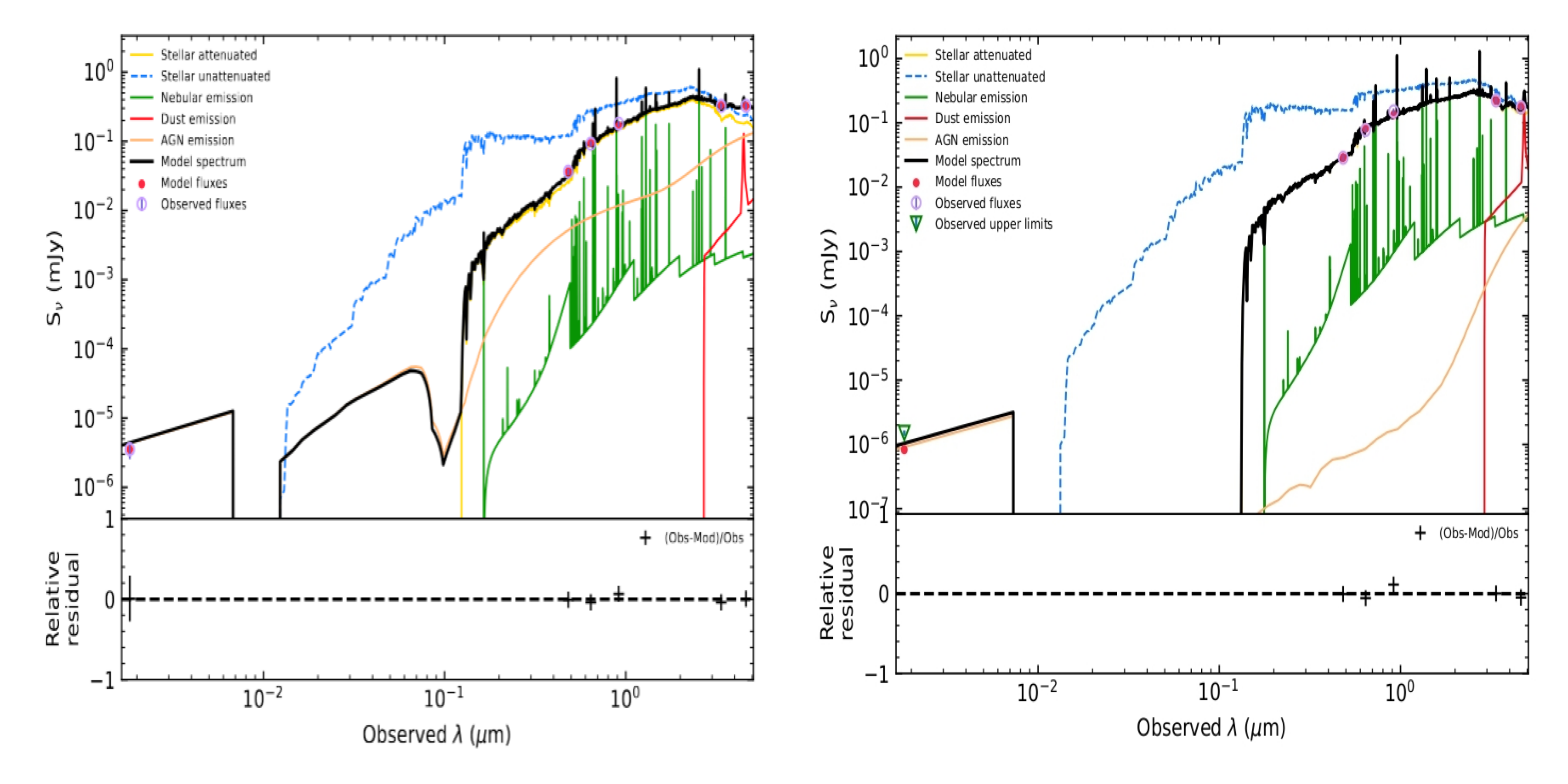}
	\caption{Example best-fitting SED models showing flux density, $S_\nu$ (mJy), as a function of observed wavelength, $\lambda$ ($\mu$m), on logarithmic scales. Individual model components are shown by coloured curves, while the total model is represented by the thick black line. Observed photometric measurements are shown as magenta circles with error bars, and the corresponding model predictions as red filled circles. The right panel includes an X-ray upper limit (green downward triangle), illustrating how non-detections are incorporated into the fitting procedure.
	}
	\label{fig:seds}
\end{figure*}

The quality of the SED fits is evaluated using the distribution of reduced $\chi^2$ values together with visual inspection of representative examples. The distribution peaks near $\chi^2_\nu \sim\!1$, with a modest tail extending toward higher values, indicating that the adopted model grid provides an adequate description of the observed photometry for most sources. Figure~\ref{fig:seds} shows two representative examples: an eFEDS detected source (left), for which the fit is constrained from the UV to the X-ray wavelengths, and an eFEDS non-detection (right), where the X-ray upper limit is shown as a downward-pointing triangle and the fit is primarily constrained by the UV, optical, and IR photometry. In both cases, the individual emission components: stellar continuum, nebular emission, dust, AGN torus, and X-ray are clearly separated, demonstrating that the adopted module configuration can describe  a range of source types.

\subsection{Derived Galaxy Properties}
\label{sec:derived_properties}

CIGALE returns posterior estimates of a range of physical parameters for each source. Here 
we focus on stellar mass, SFR, and AGN fraction, defined as the ratio of AGN to 
total IR luminosity, which form the basis of all 
subsequent analysis. We therefore examine how their distributions compare between 
the general optical population and the X-ray-detected subsample; the latter 
being, by construction, the subset where AGN activity is most unambiguously 
identified. Systematic differences between the two populations reflect both the 
physical conditions that favour detectable AGN activity and the selection function 
of eFEDS, and understanding these effects is essential before interpreting trends 
in AGN fraction as a function of stellar mass or environment.

\begin{figure}
	\centering
	\includegraphics[width=\columnwidth]{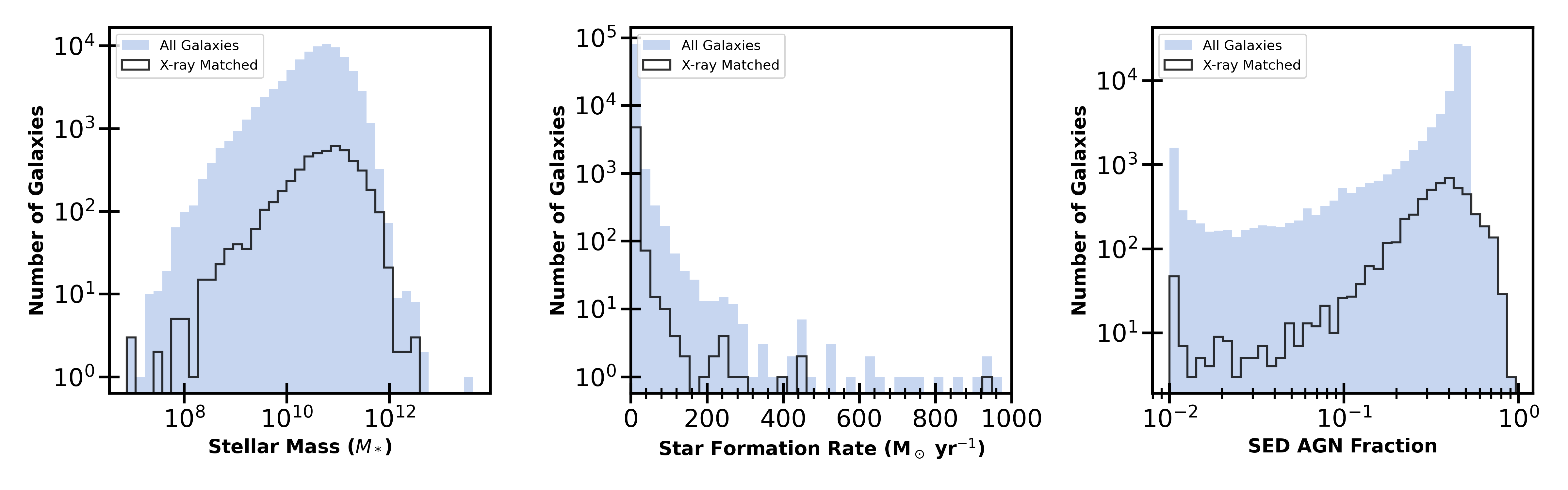}
	\caption{Comparison of galaxy and AGN properties for the full optical sample (blue) 
		and the X-ray-matched subset (black). \textbf{Left:} Stellar mass distribution ($M_{\odot}$). \textbf{Middle:} Star formation rate (SFR, $M_{\odot}~\mathrm{yr}^{-1}$). 
	    \textbf{Right:} SED-derived AGN fraction, defined as the ratio of AGN to total infrared luminosity. X-ray-detected galaxies show enhanced AGN fractions relative to the parent population, despite closely
	    tracking its stellar mass and SFR distributions. 
		} 
	\label{fig:sed_analysis}
\end{figure}

Figure~\ref{fig:sed_analysis} shows these distributions for the full optical 
sample and the X-ray-detected subsample. The AGN fraction 
distribution shows the strongest contrast: X-ray-detected galaxies exhibit 
significantly higher ratios of AGN to total IR luminosity, as expected given 
the X-ray selection. While their stellar mass and SFR distributions closely resemble those of the parent sample. This indicates that X-ray selection preferentially identifies systems with stronger AGN activity without significantly biasing the host-galaxy population.

\subsection{Galaxy Morphology}
\label{sec:morphology}

The morphological classifications provided by the Legacy Survey catalog allow 
galaxies to be grouped according to their light-profile fits. 
Table~\ref{tab:morph_type_counts} summarises the distribution across the full 
sample and the corresponding X-ray matched and stellar mass-complete subsamples. 
The sample is dominated by general S\'{e}rsic (SER) profiles, with smaller 
populations of de Vaucouleurs (DEV), exponential (EXP), and round exponential 
(REX) systems, and a minor fraction of unresolved (PSF) sources.

The X-ray matched fraction of the stellar mass-complete sample is broadly similar 
across the extended morphological classes at $4$--$7\%$, while the PSF population 
shows a noticeably elevated fraction of ${\sim}28\%$, likely reflecting 
contamination by unresolved AGN. Although morphology is not used as a primary 
analysis variable in this work, the observed distribution is consistent with the 
stellar mass and colour properties discussed in 
Sections~\ref{sec:completeness} and~\ref{sec:color_classification}.

\begin{table*}[ht]
	\centering
	\begin{tabular}{lcccc}
		\toprule
		\textbf{Morphological Type} &
		\textbf{Total Count} &
		\textbf{X-ray Matched} &
		\textbf{Mass Complete} &
		\textbf{X-ray Fraction} \\
		\midrule
		DEV (de Vaucouleurs) & 13,315 & 685  & 7,309  & 0.065 \\
		EXP (Exponential)    & 6,853  & 241  & 1,118  & 0.043 \\
		PSF (Unresolved)   & 161    & 28   & 47     & 0.277 \\
		REX (Round Exponential) & 11,962 & 502 & 1,817 & 0.067 \\
		SER (Sérsic)         & 50,502 & 2,599 & 25,750 & 0.068 \\
		\bottomrule
	\end{tabular}
	\caption{Distribution of morphological types in the DESI Legacy Survey catalogue. For each morphological class, the table lists the total number of galaxies, the number of X-ray matched sources, the number of galaxies in the stellar mass-complete sample, and the fraction of stellar mass-complete galaxies that are X-ray matched. }
	\label{tab:morph_type_counts}
\end{table*}

\subsection{Stellar Mass Completeness}
\label{sec:completeness}

In a flux-limited survey such as GAMA, progressively larger fractions of low-mass galaxies fall below the detection threshold with increasing redshift, producing an artificial turnover in the observed stellar mass function that has no physical origin. To avoid conflating this selection effect with real trends in AGN fraction or environment, we define a redshift-dependent stellar mass completeness limit above which the sample can be treated as effectively volume-limited.

We inspect the stellar mass distribution in four redshift slices over $0 < z \leq 0.4$, identifying the mass at which the distribution turns over in each bin. The resulting completeness limits are summarised in Table~\ref{tab:mstar_z_table}.

\begin{table}[h]
	\centering
	
	\begin{tabular}{c|c|c|c|c}
		\toprule
		\textbf{Redshift bin} & $z < 0.1$ & $0.1 < z \leq 0.2$ &  $0.2 < z \leq 0.3$ & $0.3 < z \leq 0.4$ \\
		$\boldsymbol{\log(M_{\star,\mathrm{lim}}/M_\odot)}$
		         & $9.0$   & $10.2$ & $10.6$  & $11.0$ \\
		\bottomrule
	\end{tabular}
	
	\caption{Redshift-dependent stellar mass completeness limits.} 
	\label{tab:mstar_z_table}
\end{table}

%
%

\begin{figure*}
	\centering
	\includegraphics[width=0.8\textwidth]{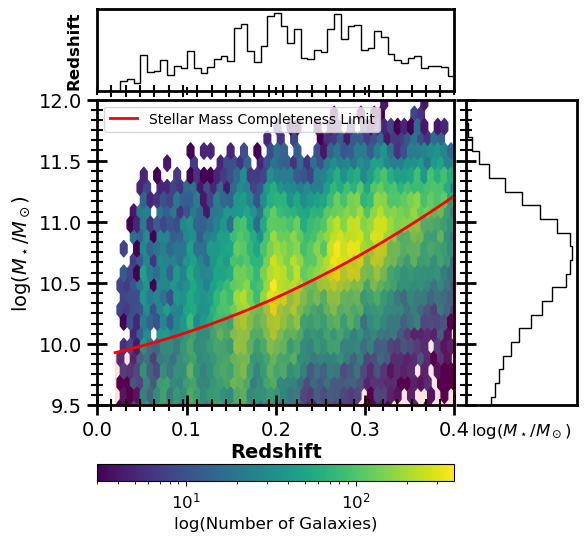}
	\caption{Stellar mass and redshift distributions of the parent sample. The 
		central panel shows the galaxy number density in the $z$--$\log(M_\star/M_\odot)$ 
		plane, visualised using hexagonal binning with logarithmic colour scaling to 
		preserve the dynamic range of the distribution. The solid red curve marks the redshift-dependent stellar mass completeness limit, above which the sample can be considered effectively volume-limited. Galaxies below this boundary are increasingly affected by the survey flux limit, leading to an artificial loss of low-mass systems at higher redshifts. The top and right panels show the corresponding marginal distributions in redshift and stellar mass, respectively.}
	\label{fig:mass_redshift}
\end{figure*}

These thresholds are indicated by the red curve in Figure~\ref{fig:mass_redshift}. All subsequent analyses of AGN fraction and environment are restricted to galaxies above the corresponding redshift-dependent limit. This cut is necessary because stellar mass is itself a strong predictor of X-ray AGN fraction; without controlling for it, apparent environmental trends may arise from the underlying mass distribution rather than from any genuine physical connection between environment and nuclear activity.

\subsection{Separating Star-forming and Quiescent Galaxies}
\label{sec:color_classification}

Rest-frame $g-i$ colour serves as our primary diagnostic for separating star-forming and quiescent galaxies, due to its sensitivity to the age of the dominant stellar population and its ability to trace the well-established bimodality of the galaxy colour distribution.
A fixed colour threshold, however, is insufficient: observed colours evolve with redshift through a combination of intrinsic stellar aging and photometric bandpass shifts, meaning a single cut would systematically misclassify galaxies across different cosmic epochs. To account for this, we adopt a redshift-dependent classification boundary following \citealt{Hang2022}

\begin{equation}
	g - i = 6.22z^2 + 1.383z + 0.831
\end{equation}

\noindent This relation was calibrated by fitting the locus of minimum number density between the red and blue sequences in the ($z$, $g-i$) plane for sources at $z<0.4$. Galaxies with colours redder than this threshold are classified as quiescent, while those bluer are designated star-forming.

This empirically motivated separation helps AGN activity to be examined independently across the two populations, and simultaneously controlling for stellar mass and large-scale environment: disentangling the different physical processes that govern each regime.


\section{AGN Identification}
\label{sec:agn_identification}

We identify AGN using four independent selection methods that probe different aspects of 
nuclear activity: X-ray emission (sensitive to both obscured and unobscured AGN), mid-IR 
colours (tracing warm dust near the nucleus), broad optical emission lines (Type~I AGN), and 
optical narrow-line ratios (BPT diagnostic). Each method has distinct selection biases and 
completeness properties, providing a means to examine how AGN demographics vary with the selection 
technique.

\subsection{X-ray AGN Selection}
\label{sec:xray_agn}

\begin{figure*}
	\centering
	\subfigure[]{
		\includegraphics[width=0.48\textwidth]{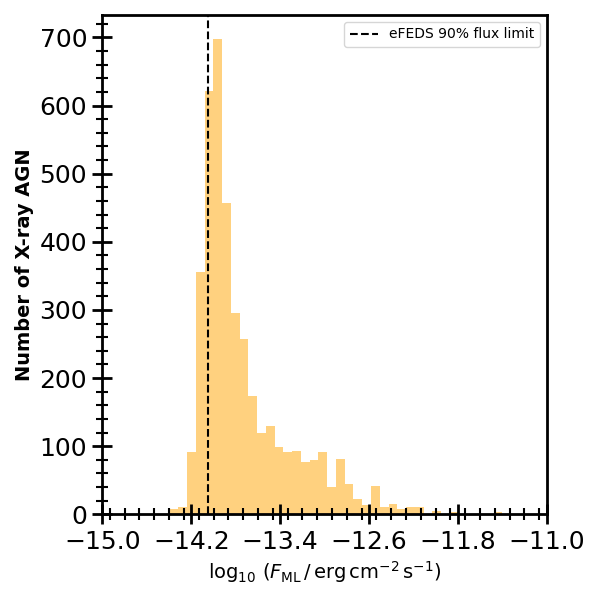}
		\label{fig:erosita_ml_flux}
	}
	\hfill
	\subfigure[]{
		\includegraphics[width=0.48\textwidth]{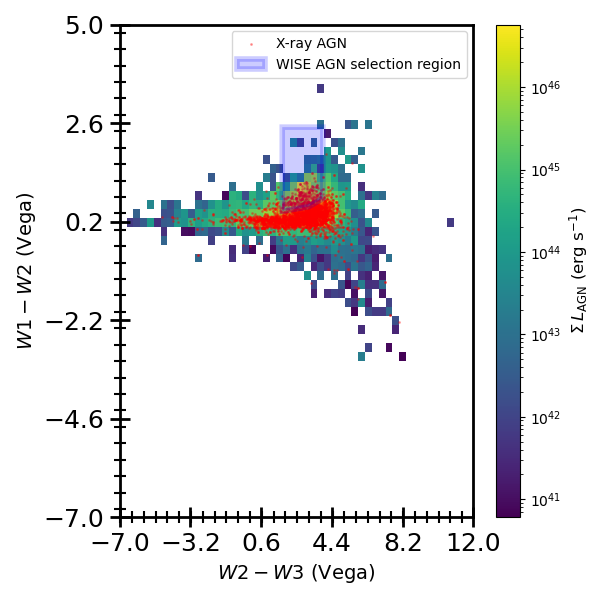}
		\label{fig:wise_cmd}
	}
	
	\vspace{0.3cm}
	
	\subfigure[]{
		\includegraphics[width=0.48\textwidth]{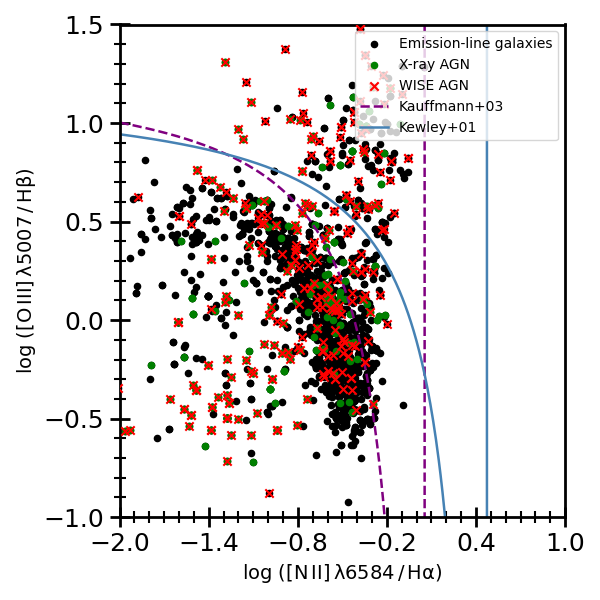}
		\label{fig:bpt}
	}
	\hfill
	\subfigure[]{
		\includegraphics[width=0.48\textwidth]{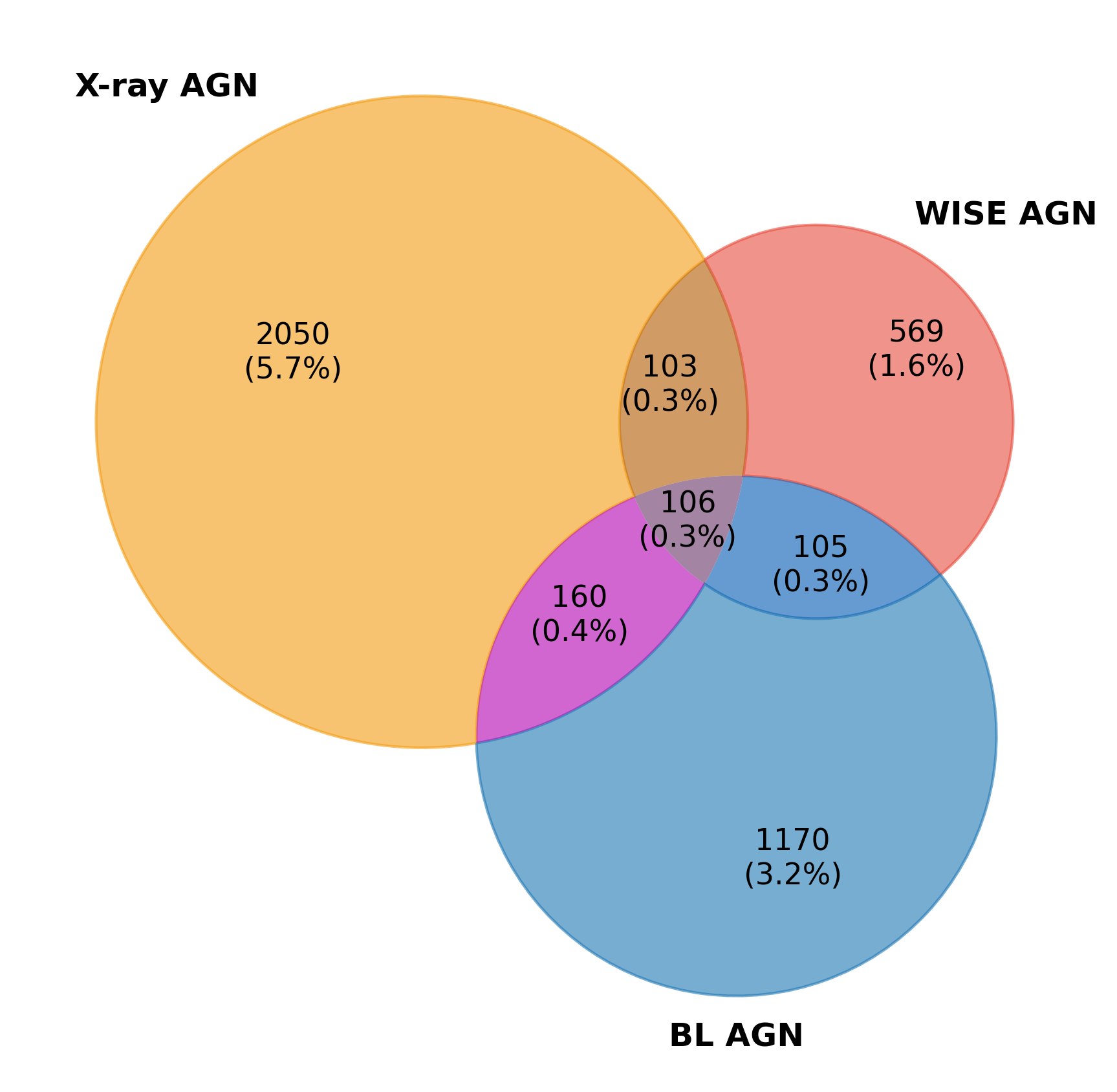}
		\label{fig:venn}
	}
	
	\caption{AGN selection methods and population overlap for the GAMA--eFEDS matched sample. \textbf{Top left:}  Distribution of observed eFEDS maximum-likelihood X-ray fluxes ($F_{\rm ML}$) for X-ray AGN. The vertical dashed line indicates the eFEDS 90\% flux completeness limit at $\log_{10}(F_{\rm ML}/{\rm erg\,cm^{-2}\,s^{-1}})=-14.05$. \textbf{Top right:} WISE colour--colour diagram ($W1-W2$ vs. $W2-W3$, Vega) for all GAMA galaxies (density map), with X-ray-selected sources above the completeness threshold overlaid as red points. The purple curve indicates the \citet{Hviding2022} selection boundary, while the colour scale represents the summed AGN luminosity. \textbf{Bottom left:} BPT diagnostic diagram of $\log\left(\frac{[\text{N\,\textsc{ii}}]\,\lambda6584}{\text{H}\alpha}\right)$ vs. $\log\left(\frac{[\text{O\,\textsc{iii}}]\,\lambda5007}{\text{H}\beta}\right)$ for emission-line galaxies in the GAMA--eFEDS matched sample with 
		$\mathrm{S/N} \geq 3$ in all four diagnostic lines at $z < 0.4$. The dashed and solid curves show the 
		empirical \citet{Kauffmann2003} and theoretical \citet{Kewley2001} 
		demarcation lines, respectively, separating star-forming galaxies, 
		composite systems, and AGN-dominated sources. Black circles show all 
		emission-line galaxies; green circles and red crosses mark X-ray and WISE AGN overlaid on the parent sample. \textbf{Bottom right:} Venn diagram illustrating the overlap among the X-ray, WISE, and BL AGN. Pairwise overlap is below $\sim\!0.4\%$ in all cases, indicating that the three methods largely probe distinct AGN populations.}

	\label{fig:agn_selection}
\end{figure*}

The top-left panel of Figure~\ref{fig:erosita_ml_flux} shows the distribution 
of maximum-likelihood X-ray fluxes, $F_{\rm ML}$, in the 0.5--2\,keV band for 
the X-ray-detected sample described in Section~\ref{sec:xrayflux}. The 
distribution peaks at $F_{\rm ML} \sim 10^{-14}$\,erg\,cm$^{-2}$\,s$^{-1}$ 
and extends to brighter fluxes with a power-law tail. For consistency with our 
stellar mass-complete sample (Section~\ref{sec:completeness}), we retain only sources 
above the eFEDS 90\% completeness limit, 
$\log(F_{\rm ML}/\mathrm{erg\,cm^{-2}\,s^{-1}}) > -14.05$, marked by the 
vertical dashed line. At the median redshift of our stellar mass-complete sample 
($\langle z \rangle = 0.2$), this flux limit corresponds to a rest-frame 
2--10\,keV luminosity of $\log(L_{\rm X}/\mathrm{erg\,s^{-1}}) \approx 42.0$, 
sufficient to identify moderate-luminosity AGN while minimising contamination 
from star-forming galaxies \citep{Ranalli2003}.

To convert observed soft-band fluxes to intrinsic rest-frame 2--10~keV luminosities, we assume 
a power-law spectrum with photon index $\Gamma = 1.8$ \citep{Nandra2007} and apply K-corrections 
using the spectroscopic redshifts from GAMA. We do not apply corrections for intrinsic absorption, 
as our goal is to identify AGN rather than to measure their intrinsic properties.

\subsection{WISE AGN Selection}
\label{sec:wise_agn}

Mid-IR colours provide an effective AGN diagnostic because hot dust in the nuclear region produces characteristically red $W1-W2$ colours (Vega magnitudes). We adopt the updated criteria 
from \citet{Hviding2022}, which identify both Type~I and Type~II AGN on the 
basis of their red mid-IR colours and achieve ${\sim}80\%$ accuracy in 
recovering spectroscopic AGN, though with some contamination from dusty 
starbursts.

The top-right panel of Figure~\ref{fig:agn_selection} shows the WISE 
colour--colour distribution of our sources, with each bin coloured by summed 
AGN luminosity. 
The selection wedge (purple curve) encloses the region containing the majority of 
high-AGN-contribution objects; sources falling within this boundary comprise 
our WISE AGN subset.

The WISE selection has the advantage of being relatively insensitive to optical obscuration, 
as mid-IR emission can penetrate moderate column densities. However, it preferentially selects 
more luminous AGN and can be contaminated by galaxies with intense star formation heating 
large amounts of dust to warm temperatures. As shown in Figure~\ref{fig:wise_cmd}, only 8.6\% of X-ray AGN fall within the WISE selection region.

\subsection{Broad-Line AGN Selection}
\label{sec:blagn}

Broad-line AGN (BL AGN) are identified by the presence of a broad component 
in their optical emission lines, reflecting high-velocity gas in the BLR that is gravitationally bound to the central SMBH. We 
identify this population by cross-matching our catalog with DESI DR1 
\citep{DESICollaboration2024, DESI2025}, which provides optical spectra covering 360--980~nm at resolutions sufficient for detailed emission-line decomposition.

\citet{Shrivastav2026} employ a line-fitting pipeline to decompose emission lines into narrow and broad components. We use these results to define our BL AGN sample, following \citet{Siudek2025}, retaining sources as BL AGN if: 
(i) the total H$\alpha$ line flux has signal-to-noise ratio (S/N) $\geq 3$; 
(ii) the broad H$\alpha$ component amplitude has S/N $\geq 3$; (iii) the ratio 
of H$\alpha$ amplitude to its uncertainty (AON) satisfies AON $\geq 3$; and 
(iv) the redshift satisfies $z \leq 0.4$, ensuring H$\alpha$ lies within the 
observed spectral coverage. By construction, all sources satisfying these 
criteria have detectable emission-line widths suitable for virial black hole 
mass estimation (Section~\ref{sec:bh_mass}).

This selection is highly specific but limited in completeness, as it requires 
both an unobscured view of the BLR and sufficient S/N in the broad-line 
component. The resulting sample is therefore biased toward Type~I (unobscured) 
systems with higher accretion rates and represents a subset of the total AGN 
population.

\subsection{BPT AGN Selection}
\label{sec:bpt_agn}

We classify narrow-line galaxies using the Baldwin--Phillips--Terlevich (BPT) diagnostic 
diagram \citep{Baldwin1981}, which separates star-forming systems from AGN by exploiting the harder ionising spectrum of accreting black holes 
relative to young, massive stars. The diagram plots the logarithmic ratios 
of forbidden to Balmer lines

\begin{equation}
	\log\left(\frac{[\text{N\,\textsc{ii}}]\,\lambda6584}{\text{H}\alpha}\right) 
	\quad\text{vs.}\quad 
	\log\left(\frac{[\text{O\,\textsc{iii}}]\,\lambda5007}{\text{H}\beta}\right),
	\label{eq:bpt}
\end{equation}

\noindent requiring S/N~$\geq 3$ in all four lines to ensure reliable placement in the diagram.

The bottom-left panel of Figure~\ref{fig:bpt} shows the BPT diagnostic diagram for galaxies in our spectroscopic sample. 
The standard classification curves from \citet{Kauffmann2003} (purple dashed line) and 
\citet{Kewley2001} (red solid line), are overplotted. Galaxies below the Kauffmann line are 
classified as star-forming (BPT SF), while objects above the Kewley line are identified as 
AGN (BPT AGN). Systems falling in the region between the two boundaries occupy the composite 
zone, where both star formation and nuclear activity contribute to the observed line excitation.

The BPT selection is most sensitive to AGN with strong narrow-line emission and is relatively 
insensitive to obscuration of the continuum source. However, it can miss low-ionisation AGN 
and systems where star formation dilutes the AGN emission-line signature. Additionally, the 
BPT diagram requires detection of $[\text{O\,\textsc{iii}}]$ and H$\beta$, which may be weak or absent in some 
AGN, notably those in early-type hosts with little ongoing star formation. We also locate X-ray and WISE AGN on the BPT diagram and find that 
the BPT diagnostic has limited discriminating power for these populations.

\subsection{Black Hole Mass Estimation}
\label{sec:bh_mass}

To quantify the growth of SMBHs in our AGN sample, we estimate 
black hole masses using the virial method based on single-epoch spectroscopy. This method 
relies on the assumption that gas in the BLR is virialized under the 
gravitational influence of the central SMBH. The black hole mass can be expressed as

\begin{equation}
	M_{\rm BH} = f\,\frac{R_{\rm BLR}\,(\mathrm{FWHM})^2}{G},
	\label{eq:mbh_virial}
\end{equation}

\noindent where $R_{\rm BLR}$ is the BLR radius, FWHM is the full width at half maximum of the line profile, $G$ is the gravitational constant, and $f$ is a scaling factor accounting for the geometry and kinematics of the BLR. We relate $R_{\rm BLR}$ to the line luminosity via the radius--luminosity relation \citep{Kaspi2000, Bentz2009}, applying the calibrations of 
\citet{Greene2005} for H$\alpha$ and \citet{Vestergaard2006} for H$\beta$. When both lines are available, we 
prefer the H$\alpha$-based estimate given its typically higher S/N.

The resulting mass estimates allow us to compute Eddington ratios
\begin{equation}
	\lambda_{\rm Edd} = \frac{L_{\rm bol}}{L_{\rm Edd}} = 
	\frac{L_{\rm bol}}{1.26 \times 10^{38}(M_{\rm BH}/M_\odot)~
		\mathrm{erg~s^{-1}}},
	\label{eq:eddington_ratio}
\end{equation}
where $L_{\rm bol}$ is the AGN bolometric luminosity and $L_{\rm Edd}$ 
assumes a fully ionised hydrogen plasma with electron-scattering opacity. 
For X-ray-detected AGN, $L_{\rm bol}$ is estimated from the X-ray luminosity 
using the bolometric correction of \citet{Lusso2012}; for BL~AGN, we use 
the H$\alpha$ luminosity with the correction of \citet{Greene2005}.

These Eddington ratios provide a normalized measure of the accretion rate, 
enabling us to test whether environment modulates not just AGN incidence 
but also the underlying accretion physics.

\subsection{Summary of AGN Samples}
\label{sec:agn_summary}
Table~\ref{tab:agn_summary} summarizes the four AGN selection methods, 
their selection criteria, and key biases. The negligible overlap between 
samples ($\leq 0.4\%$ for any pair) reflects the diverse physical conditions 
under which AGN can be detected, and motivates our approach of examining 
environmental trends independently for each selection; a comparison that 
reveals whether any such trends are universal or method-dependent.


\begin{table*}[ht]
    \centering
    \setlength{\tabcolsep}{1pt}
    \begin{tabular}{c@{\hspace{-6pt}}c@{\hspace{-6pt}}c@{\hspace{-6pt}}c}
        \toprule
        \textbf{Method} &
        \makecell{\textbf{Selection}\\\textbf{Criterion}} &
        \makecell{\textbf{Key}\\\textbf{Advantage}} &
        \makecell{\textbf{Key Biases/}\\\textbf{Limitations}} \\
        \midrule

        X-ray &
        \makecell{$\log(F_{\rm ML}) > -14.05$} &
        \makecell{Detects obscured and\\
        unobscured AGN; relatively\\
        unbiased} &
        \makecell{Flux-limited; misses heavily\\
        obscured Compton-thick AGN} \\

        \midrule

        WISE &
        \makecell{Colour cuts in $W1-W2$\\
        vs. $W2-W3$} &
        \makecell{Penetrates moderate obscuration;\\
        large sky coverage} &
        \makecell{Contamination from dusty starbursts;\\
        biased toward luminous AGN} \\

        \midrule

        BL AGN &
        \makecell{Broad H$\alpha$ with S/N $\geq 3$,\\
        AON $\geq 3$} &
        \makecell{Direct BH mass estimates;\\
        unambiguous AGN identification} &
        \makecell{Only Type I; requires high S/N;\\
        orientation dependent} \\

        \midrule

        BPT &
        \makecell{Above Kewley line in\\
        $\left[\mathrm{O\,III}\right]/\mathrm{H\beta}$ vs.\\
        $\left[\mathrm{N\,II}\right]/\mathrm{H\alpha}$} &
        \makecell{More complete for luminous\\
        narrow-line (Type II) AGN} &
        \makecell{Misses low-ionisation AGN;\\
        contamination from shocks/LINERs} \\

        \bottomrule
    \end{tabular}

    \caption{Summary of AGN selection methods. Each approach has distinct selection biases and completeness properties, enabling us to examine demographic variations across selection techniques and assess the robustness of environmental trends among different source populations.}

    \label{tab:agn_summary}
\end{table*}


\section{Environmental Measurements}
\label{sec:environment}

We characterise the large-scale environment of each galaxy in the GAMA G09 
sample following the tidal tensor formalism of \citet{Alam2018}. Galaxy 
positions are embedded in a three-dimensional Cartesian coordinate system,  and the local number density field is estimated via Voronoi tessellation \citep{vandeWeygaert2012}, with the GAMA random catalog used to account for survey boundaries and masked regions. The resulting density field is smoothed with an isotropic Gaussian kernel of 
width $5\,h^{-1}\,\mathrm{Mpc}$ to produce the smoothed galaxy over-density 
$\delta_{\mathrm{g}}^{\mathrm{s}}$, which is related to a dimensionless 
gravitational potential $\tilde{\Phi}$ through the Poisson equation 
$\nabla^2\tilde{\Phi} = \delta_{\mathrm{g}}^{\mathrm{s}}$. 

The local galaxy overdensity within a sphere of radius $R$ is defined as
\begin{equation}
	\delta_{R} = \frac{n_{g} - \langle n_{g} \rangle}{\langle n_{g} \rangle},
	\label{eq:delta}
\end{equation}
where $n_{g}$ is the local galaxy number density and $\langle n_{g} \rangle$ is its
mean.  We adopt $R = 5\,h^{-1}$\,Mpc for the local over-density, but also look at R dependence to study the robustness of features.

\subsection{The $\delta_{\rm rank}$ Methodology}
\label{sec:delta_rank}

While $\delta_{R}$  provide a rich characterisation of the cosmic
web, applying them directly to study AGN--environment correlations faces a
fundamental complication: more massive galaxies preferentially reside in denser
environments \citep{Peng2010, Darvish2016}. Since the AGN fraction itself rises
steeply with stellar mass, any naive correlation between AGN incidence and
$\delta_{R}$ risks being driven entirely by this underlying mass--environment
relation rather than by a genuine environmental influence on nuclear activity.
Previous studies have attempted to control for this by simultaneously binning in
stellar mass and environment, but this approach is limited by poor statistics in
individual bins and does not fully remove residual correlations when the bins are
broad. To circumvent this problem cleanly, we introduce the $\delta_{\rm rank}$
parameter, which measures the relative overdensity of each galaxy compared to
other galaxies of similar stellar mass. This is similar to rank based environment definition adopted in \citet{2021MNRAS.502.3242X,2024MNRAS.527.3771A} .The procedure is as follows:

\begin{enumerate}
	\item Divide the stellar mass-complete sample into narrow stellar mass bins of
	width $\Delta \log M_{\star} = 0.2$~dex.
	\item Within each bin, rank galaxies in ascending order of their $\delta_{R}$
	value.
	\item Assign each galaxy a normalized rank,
	\begin{equation}
		\delta_{\rm rank} = \frac{{\rm rank} - 1}{N_{\rm bin} - 1},
		\label{eq:rank}
	\end{equation}
	where $N_{\rm bin}$ is the number of galaxies in that mass bin, so that
	$\delta_{\rm rank} \in [0,1]$.
	\item Combine galaxies from all mass bins for subsequent analysis.
\end{enumerate}

\begin{figure}
	\centering
	\includegraphics[width=\textwidth]{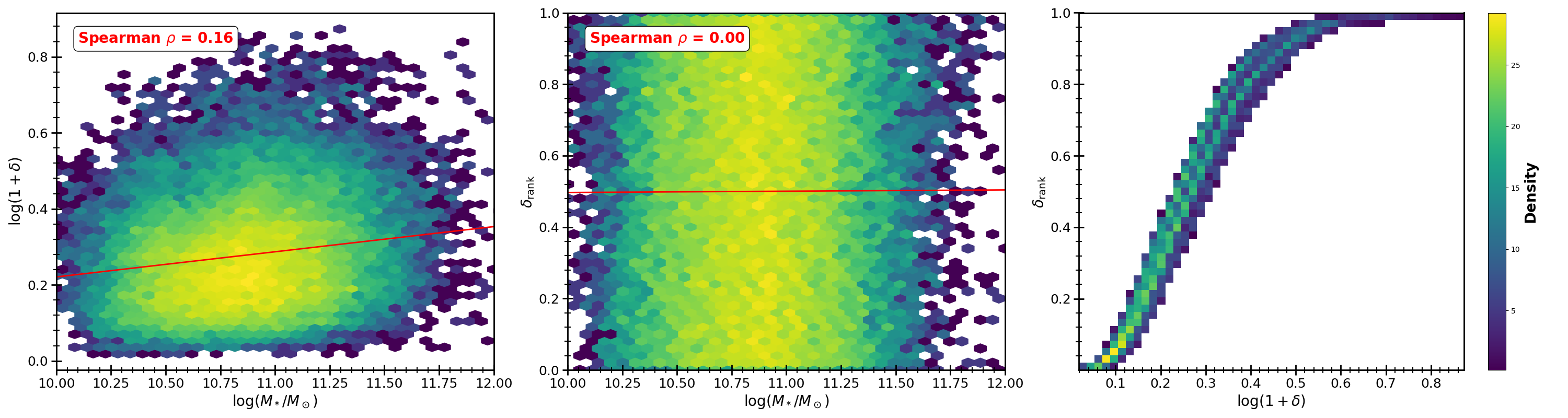}
	\caption{Dependence of the local environment on stellar mass for the stellar mass-complete sample. 
		\textbf{Left:} Hexagonal density map of $\log(1+\delta)$ as a function of stellar mass, where $\delta$ is the galaxy overdensity measured within an $5\,h^{-1}\,\mathrm{Mpc}$ aperture. 
		The solid red line shows the best-fitting linear relation, and the quoted value indicates the Spearman rank correlation coefficient. 
		\textbf{Middle:} Same as the left panel, but using the rank-normalised overdensity parameter $\delta_{\rm rank}$, which removes sensitivity to the absolute scale of the density field; the corresponding Spearman coefficient is consistent with zero by
		construction.
		\textbf{Right:} Hexagonal density map of the rank-normalised overdensity parameter $\delta_{\rm rank}$ as a function of $\log(1+\delta)$ for the stellar mass-complete sample. The monotonic mapping reflects the rank-normalisation procedure, which removes sensitivity to the absolute scale of the density field. 
	}
	\label{fig:delta_rank_mstar}
\end{figure}

By construction, the resulting $\delta_{\rm rank}$ distribution is uniform across
all stellar mass bins, and its correlation with stellar mass is zero. Any trend
observed as a function of $\delta_{\rm rank}$ therefore cannot be attributed to the
mass--environment degeneracy and must reflect a genuine environmental dependence.
Figure~\ref{fig:delta_rank_mstar} demonstrates the effectiveness of our approach: the left
panel shows a positive correlation between $\delta_{R}$ and stellar mass in our
sample (Spearman $\rho \approx 0.16$), while the middle panel shows that
$\delta_{\rm rank}$ has zero correlation with stellar mass by construction
(Spearman $\rho \approx 0.0$). This allows us to cleanly isolate environmental
effects.

It is important to understand what $\delta_{\rm rank}$ represents physically.
Unlike $\delta_{R}$, which measures absolute over-density at a fixed physical scale,
$\delta_{\rm rank}$ measures the relative density environment of a galaxy compared
to other galaxies of similar mass. A galaxy with $\delta_{\rm rank} = 0.1$ resides
in the bottom 10\% of environments for its mass, regardless of whether it is a
low-mass or high-mass galaxy.
This relative ranking allows a fair comparison across the full
stellar mass range without requiring separate mass bins. However, this approach
also means that $\delta_{\rm rank}$ does not directly correspond to a fixed
physical scale or over-density. A galaxy with $\delta_{\rm rank} = 0.9$ may reside
in a group at $M_{\star} = 10^{10}\,M_{\odot}$ but in a cluster at
$M_{\star} = 10^{11}\,M_{\odot}$, since more massive galaxies on average inhabit
denser regions. Therefore, when interpreting results as a function of
$\delta_{\rm rank}$, we are asking: ``Does environment matter for AGN activity
after accounting for stellar mass?''

\section{Results}
\label{sec:results}
We first note that the three AGN selection methods yield somewhat different AGN fractions within the sample. The X-ray, WISE, and BL selections identify AGN fractions of 6.7\%, 2.5\%, and 4.3\%, respectively. Although the differences in the individual fractions are modest, a more striking result is the remarkably small overlap between the selected populations. As shown in Figure~\ref{fig:agn_selection}, the pairwise overlap between any two methods is only $\sim$ 0.3--0.4\%, indicating that the vast majority of AGN identified by one diagnostic are not recovered by the others. This suggests that each method preferentially traces different manifestations of AGN activity rather than a common underlying population.

Consequently, combining all three diagnostics increases the total AGN fraction to approximately 12\%. The limitations of single-waveband AGN selection, and the physical origin of the low overlap between different diagnostics, are reviewed extensively by \citet{Hickox2018}. Given that the three selection methods are sensitive to distinct aspects of AGN activity, any environmental dependence may likewise differ between AGN classes. This motivates a separate investigation of environmental trends for each selection method, which we undertake in the remainder of this section.

\subsection{Stellar Mass Dependence of AGN Fraction}
\label{sec:stellar_mass}

\begin{figure}
	\centering
	\includegraphics[width=\columnwidth]{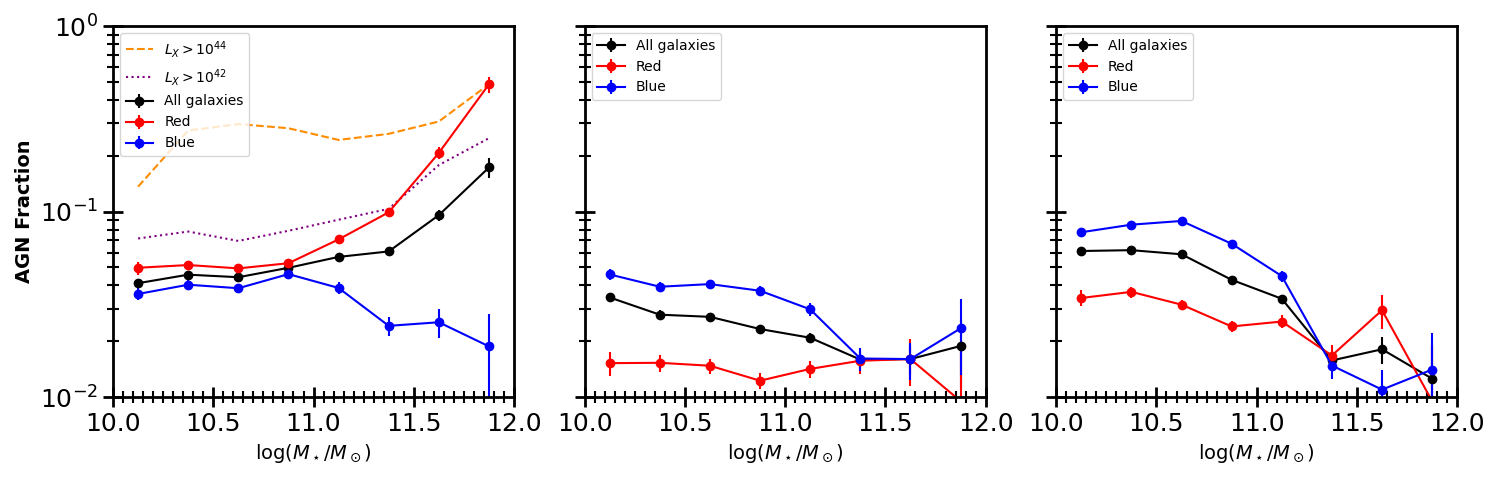}
	\caption{AGN fraction as a function of stellar mass ($M_\star$) for the 
		matched sample. In each panel, black, red, and blue points correspond to 
		the full, quiescent, and star-forming galaxy subsamples, respectively.
		\textbf{Left:} X-ray AGN fraction. Dashed and dotted curves show results 
		above luminosity thresholds of $L_X > 10^{44}$ and 
		$L_X > 10^{42}\,\mathrm{erg\,s^{-1}}$, respectively.
		\textbf{Middle:} WISE AGN fraction.
		\textbf{Right:} BL AGN fraction.
		Error bars represent binomial uncertainties; bins with fewer than ten 
		galaxies are omitted.}
	\label{fig:agn_frac_mstar}
\end{figure}

We first characterise the well-established dependence of AGN fraction on
stellar mass ($M_\star$). Figure~\ref{fig:agn_frac_mstar} shows the AGN fraction as
a function of stellar mass for the three selection methods: X-ray, WISE,
and BL AGN, displayed in the left, middle, and right panels,
respectively. In each panel, the black, red, and blue points correspond to the
full sample, quiescent, and star-forming galaxies, respectively.

The X-ray AGN fraction rises from $\sim\!4\%$ to $\sim\!15\%$ across the
stellar mass range for the full sample, but the red galaxies show a
stronger increase of AGN fraction all the way to $\approx\!40\%$ at
$M_\star \sim 10^{12}\,M_\odot$. This enhancement among massive passive hosts
is connected to the broadly flat or mildly declining X-ray AGN fraction seen in
star-forming galaxies at high stellar masses. \citet{Aird2012_primus} show a similar trend, though they attribute the rise primarily to
selection effects rather than a genuine increase in AGN activity.

The WISE AGN fraction slightly decreases with the stellar mass for the full sample. 
For the blue star-forming galaxies, this fraction shows a strong decrease above $M_\star \sim 10^{11}\,M_\odot$, a trend aligning with studies demonstrating that 
mid-IR colour selections ($W1-W2 \geq 0.8$) isolate dust-obscured, high-accretion AGN which naturally decline as massive star-forming hosts undergo 
rapid quenching \citep{2016ApJ...832..119H, 2024ApJ...963...53C}. In contrast, for the red quiescent 
galaxies, the active fraction exhibits no significant dependence on the stellar mass. This 
mass-independent flat trend indicates a distinct population of low-luminosity or radio-mode 
AGN in passive hosts, whose maintenance-mode triggering mechanisms do not scale strictly with 
the cold-gas reservoirs or the cumulative mass of the remaining host stellar population \citep{Best2005}.

The BL AGN active fraction is higher overall than that of mid-IR selections, hovering near 8\% at the low-mass end before dropping sharply above $M_\star \sim 10^{11}\,M_\odot$ to approximately 1\%. This declining mass trend remains fundamentally similar across the red, blue, and full galaxy populations. However, a clear  difference is visible in the low-mass regime, where the BL AGN
fraction in star-forming galaxies exceeds that in quiescent hosts by a factor of $\sim\!3$. This prominent excess in gas-rich environments is consistent with the standard paradigm that high-accretion, unobscured Type I AGN are preferentially driven by young, star-forming stellar populations where ample cold gas is readily funneled to the central SMBH \citep{Kauffmann2003, Jin2026}. As host galaxies evolve past the $10^{11} M_\odot$ threshold, powerful feedback or rapid gas consumption triggers quenching, causing the drop in the active fraction of blue hosts and leaving behind a residual population of low-efficiency accretion modes in massive, early-type galaxies \citep{Pimbblet2013, Jin2026}.

\subsection{Environmental Dependence of AGN Fraction}
\label{sec:env_dependence}

\begin{figure*}
	\centering
	\includegraphics[width=\textwidth]{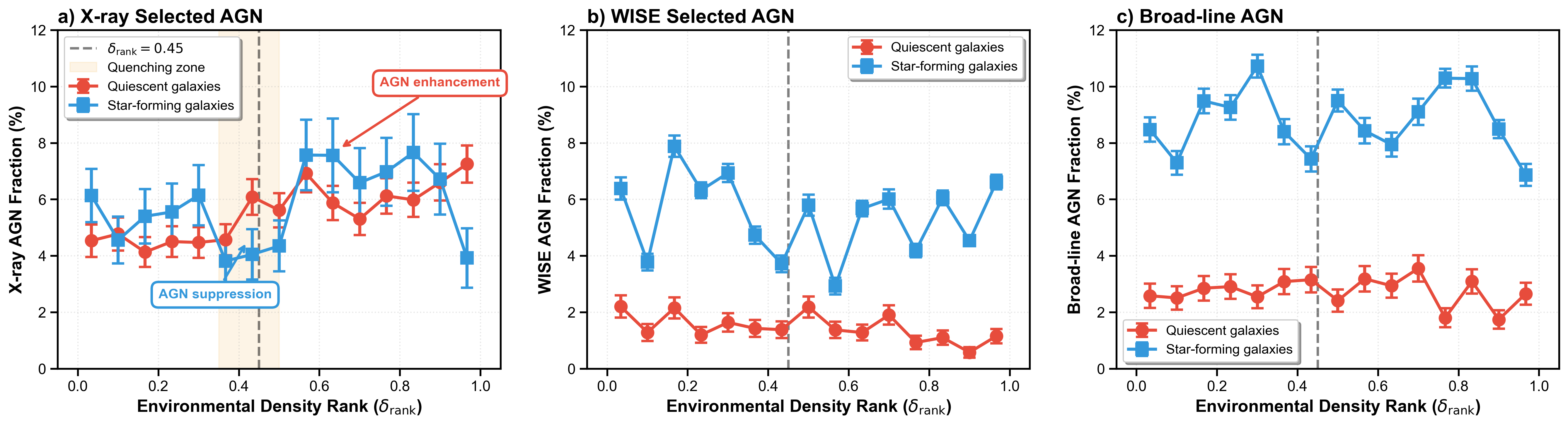}
	\caption{AGN fraction as a function of environmental density rank ($\delta_{\rm rank}$) for the three AGN selection methods, shown separately for quiescent (red) and star-forming (blue) galaxies. The X-ray-selected sample shows a distinct transition near $\delta_{\rm rank}\approx0.4$, marked by a sharp increase in the AGN fraction and highlighted by the shaded region. In contrast, WISE AGN show no significant dependence on environment, BL AGN show largely flat trends with only minor variations. Error bars represent binomial uncertainties.}
	\label{fig:agn_delta_rank}
\end{figure*}

Figure~\ref{fig:agn_delta_rank} presents the AGN fraction as a function of $\delta_{\rm rank}$ for the three AGN selection methods, separated by galaxy colour.

For X-ray AGN, we observe a clear step-function behavior in red galaxy populations. The AGN fraction remains approximately constant at $\sim$ 4\% for $\delta_{\rm rank} < 0.35$, then transitions sharply to $\sim$ 6\% for $\delta_{\rm rank} > 0.5$, representing a $\sim$ 50\% increase in AGN incidence in high-density environments. This transition is remarkably sharp, occurring over $\Delta \delta_{\rm rank} \approx 0.15$. Blue galaxies display 
a broadly similar low-to-high density trend, but show a dip at intermediate 
$\delta_{\rm rank}$ not present in the red population, suggesting an additional 
triggering channel in star-forming hosts.

In stark contrast, WISE AGN show no significant environmental trend, 
maintaining roughly constant fractions of $\sim\!6\%$ and $\sim\!2\%$ for 
blue and red galaxies, respectively, across all $\delta_{\rm rank}$ values. 
BL AGN similarly exhibit flat or weakly declining fractions with 
$\delta_{\rm rank}$, at $\sim\!8\%$ and $\sim\!3\%$ for blue and red 
galaxies. The absence of environmental dependence in these optical/IR-selected samples suggests that different selection methods are sensitive to distinct physical mechanisms or AGN populations.

\subsection{AGN Fraction as a Function of SFR}
\label{sec:sfr_dependence}

\begin{figure*}
	\centering
	\includegraphics[width=\textwidth]{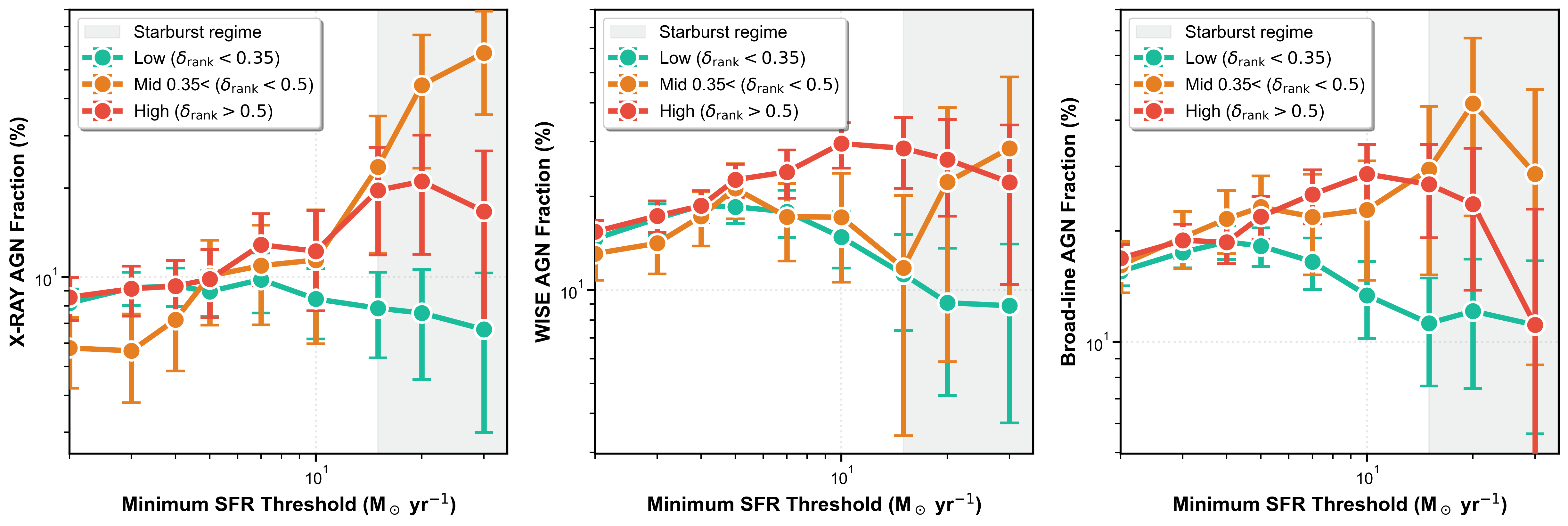}
	\caption{AGN fraction as a function of SFR threshold for the three AGN selection methods, shown in three environmental-density bins: low ($\delta_{\rm rank}<0.35$), intermediate ($0.35<\delta_{\rm rank}<0.5$), and high ($\delta_{\rm rank}>0.5$). A strong dependence on star formation activity is evident in denser environments, especially for X-ray AGN, whose fraction increases sharply toward the most actively star-forming galaxies. WISE AGN show a more gradual rise with SFR, whereas BL AGN show only a weak dependence. The shaded region marks the starburst regime. Error bars represent binomial uncertainties.}
	\label{fig:agn_sfr_environment}
\end{figure*}

To further investigate the interplay between star formation, environment, and AGN activity, we examine the AGN fraction as a function of SFR in three $\delta_{\rm rank}$ bins: low ($\delta_{\rm rank} < 0.35$), intermediate ($0.35 < \delta_{\rm rank} < 0.5$), and high ($\delta_{\rm rank} > 0.5$). Results are shown in Figure~\ref{fig:agn_sfr_environment}.

In low-density environments ($\delta_{\rm rank} < 0.35$), all three selection methods show modest decrease in AGN fraction with increasing star formation activity, remaining below 10\%  even for the most actively star-forming galaxies, with broadly similar fractions across methods.

At intermediate densities ($0.35 < \delta_{\rm rank} < 0.5$), striking differences emerge between selection methods. X-ray AGN show a dramatic rise, reaching $\sim$ 60\% for galaxies with SFR $> 15~M_{\odot}~{\rm yr}^{-1}$. WISE-selected AGN track the low-density trend up to SFR $\sim 5~M_{\odot}~{\rm yr}^{-1}$, then transition to fractions comparable to high-density regions. BL AGN fractions in this density range closely follow the high-density trend across the full SFR range.

In high-density environments ($\delta_{\rm rank} > 0.5$), all three methods show rising AGN fractions toward high SFRs, converging at $\sim\!20$--$25\%$.
These results suggest that high-density environments preferentially enhance AGN triggering in star-forming galaxies, with X-ray selection showing the strongest environmental response and WISE and BL AGN exhibiting more modest enhancement.

\begin{figure*}
	\centering
	\includegraphics[width=0.6\textwidth]{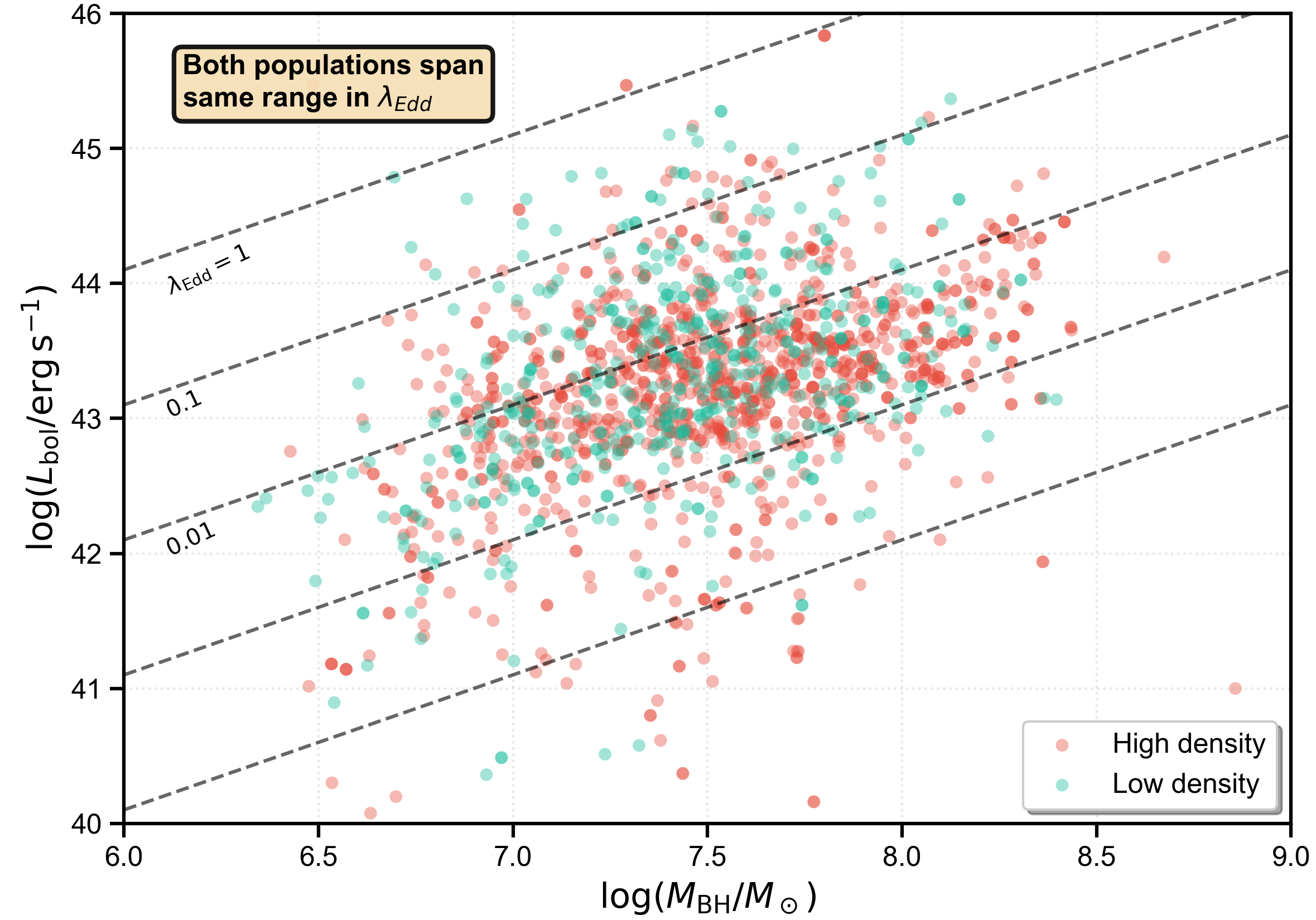}
	\caption{Eddington diagram showing 
		bolometric luminosity ($\log L_{\rm bol}$/erg\,s$^{-1}$) vs. black hole mass 
		($\log M_{\rm BH}/M_\odot$), with lines of constant $\lambda_{\rm Edd}$. 
		AGN in both environments occupy comparable regions of the parameter space, 
		spanning similar ranges in accretion rate and black hole mass. This 
		suggest that environment influences the incidence of AGN activity 
		rather than the underlying accretion properties of the active population.} 
	\label{fig:eddington_distribution}
\end{figure*}

To investigate whether environment affects not just the AGN fraction but also the accretion physics, we look at the eddington diagram for the X-ray AGN in low and high density (Figure~\ref{fig:eddington_distribution}). We find no measurable difference in the bolometric luminosity ($\log L_{\rm bol}$/erg\,s$^{-1}$) vs. black hole mass 
($\log M_{\rm BH}/M_\odot$) space for different density bins. 
This result indicates that while environment modulates the AGN fraction, it does not significantly affect the accretion physics once an AGN is triggered. The increased AGN fraction in high-density regions appears to reflect a higher triggering efficiency rather than a change in the typical accretion rate.

\subsection{Robustness Tests: Overdensity Scale Dependence}
\label{sec:robustness}

\begin{figure*}
	\centering
	\includegraphics[width=\textwidth]{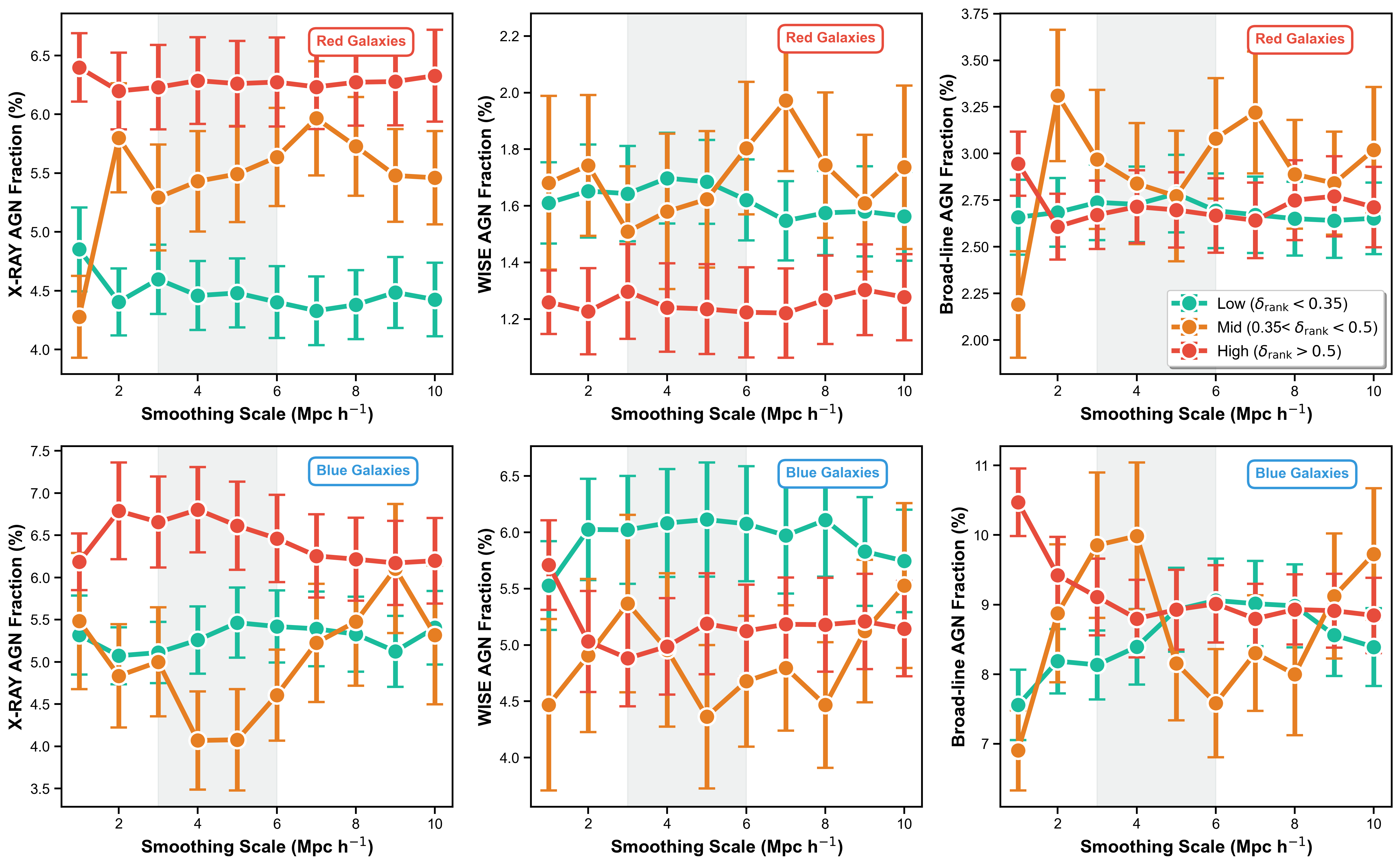}
    \caption{AGN fraction as a function of environmental smoothing scale for X-ray, WISE, and BL AGN, shown separately for red (top) and blue (bottom) galaxies in three environmental-density regimes. The environmental dependence of X-ray AGN remains robust on intermediate scales, with the strongest contrast observed at $\sim5\,h^{-1}\,\mathrm{Mpc}$. At smaller smoothing scales the trends become increasingly noisy, while at larger scales the environmental signal weakens. In contrast, WISE and BL AGN show little dependence on smoothing scale. The shaded region marks the range of scales over which the environmental effect is most pronounced. Error bars represent binomial uncertainties.}
	\label{fig:scale_dependence}
\end{figure*}

To test the robustness of our results and identify the physical scale most relevant for AGN-environment correlations, we repeat our analysis using overdensities measured at different scales between 1-10 $\,h^{-1}$\,Mpc (Figure~\ref{fig:scale_dependence}).

The environmental dependence of the X-ray AGN fraction remains qualitatively 
stable between 3 and $6\,h^{-1}$\,Mpc. The step-function behavior is most 
pronounced around $5\,h^{-1}$\,Mpc, with slightly weaker but still significant 
trends at 3--$6\,h^{-1}$\,Mpc. At $2\,h^{-1}$\,Mpc the signal becomes 
noisier due to reduced number statistics, though the overall trend persists. 
Beyond $6\,h^{-1}$\,Mpc the correlation weakens considerably. Together, 
these results imply that the dominant environmental scale for X-ray AGN 
triggering is $\sim\!5\,h^{-1}$\,Mpc, corresponding to intermediate-
density structures such as filaments, sheets, but  much larger than group-scale environments.

\subsection{Physical Interpretation of the X-ray Environmental Dependence}

The step-function transition in the X-ray AGN fraction at 
$\delta_{\mathrm{rank}} \sim 0.4$, rising from $\sim\!4\%$ in underdense 
regions to $\sim\!6\%$ at intermediate overdensities, points toward large-scale 
environmental processes operating at the boundaries of the cosmic web. This 
interpretation is supported by the scale-dependence analysis of 
Section~\ref{sec:robustness}, where the signal peaks at smoothing scales of 
$\sim\!5\,h^{-1}\,\mathrm{Mpc}$ and weakens beyond. Since these scales greatly 
exceed the virial radii of individual clusters, a localized cluster-core 
mechanism is disfavoured; instead, the signal implicates the intermediate-density 
topology of the cosmic web, namely filaments, sheets, and infalling 
envelopes surrounding massive structures \citep{Koulouridis2024}.

Within these large-scale structural pathways, galaxies undergo substantial pre-processing before entering dense node environments. The moderate overdensities found within $\sim\!5h^{-1}\,\mathrm{Mpc}$ of filaments facilitate a high frequency of galaxy--galaxy tidal interactions and stochastic cold-gas accretion streamed directly from the intergalactic medium \citep{Umehata2019, Bulichi2024}. Crucially, these large-scale environment-driven interactions can efficiently funnel cold gas reservoirs toward nuclear regions without necessarily triggering a burst of widespread star formation. This provides a natural physical explanation for why the observed X-ray AGN enhancement is prominent across both the red quiescent and blue star-forming populations, rather than being exclusive to the active main sequence. Consequently, the transition at $\delta_{\text{rank}} \sim 0.4$ likely maps the physical boundary where galaxies transition from the isolated, stochastic accretion regimes of cosmic voids into the dynamically active, fuel-rich corridors of large-scale structure.

The absence of a comparable environmental signal in mid-IR and 
broad-line selected AGN is supporting the picture that those selections 
preferentially recover more luminous or unobscured systems whose triggering 
is less sensitive to the large-scale tidal field probed here, and may 
additionally be affected by orientation-dependent obscuration biases in 
denser environments.

\section{Summary and Conclusions}
\label{sec:conclusions}

We have conducted a comprehensive multi-wavelength study of AGN activity in the GAMA-eFEDS overlap region, examining  galaxies with stellar masses, SFRs, and AGN identifications from X-ray, optical, and IR data. Our principal findings are as follows.

\begin{enumerate}
	
	\item \textbf{Novel methodology for isolating environmental effects:} We introduce the $\delta_{\rm rank}$ parameter, which ranks galaxies by local overdensity within narrow stellar mass bins. This technique effectively decouples stellar mass and environmental effects, enabling a clean measurement of genuine environmental dependencies while mitigating selection effects.
	
	\item \textbf{Clear environmental dependence of X-ray AGN:} X-ray AGN exhibit a significant step-function increase in AGN fraction with environment, rising from $\sim\!4\%$ in low-density regions to $\sim\!6\%$ in high-density environments, with a sharp transition at $\delta_{\rm rank}\sim\!0.4$. This $\sim\!50\%$ enhancement is observed in both blue and red galaxy populations, demonstrating that the environmental effect is independent of current star formation activity or galaxy colour.
	
	\item \textbf{Selection method matters:} WISE and BL AGN show no significant environmental dependence, maintaining constant AGN fractions across all density regimes. This stark contrast with X-ray results suggests that different selection methods probe distinct AGN populations or physical mechanisms, with X-ray selection being most sensitive to environmentally-triggered AGN activity.
	
	\item \textbf{SFR-environment interplay reveals complex triggering:} The relationship between AGN fraction and SFR varies dramatically with both environment and selection method. In high-density regions, X-ray AGN fractions reach $\sim\!20\%$ for galaxies with $\mathrm{SFR} > 15\,M_\odot\,\mathrm{yr}^{-1}$, compared to $<\!10\%$ in low-density regions. The intermediate-density regime shows peculiar behavior, with X-ray AGN fractions reaching $\sim\!60\%$ at high SFRs, while WISE and BL AGN selections display distinct evolutionary paths. This suggests multiple pathways for AGN triggering depending on both environment and star formation activity.
	
	\item \textbf{Environment modulates triggering, not accretion:} The ($\log L_{\rm bol}$/erg\,s$^{-1}$) vs. black hole mass 
	($\log M_{\rm BH}/M_\odot$) space shows no appreciable variation across different density environments. This indicates that while high-density environments enhance AGN triggering efficiency, they do not fundamentally alter the accretion physics or typical accretion rates once an AGN is active.
	
	\item \textbf{Relevant physical scales:} Our robustness tests reveal that the environmental dependence is strongest at scales of $\sim\!5\,h^{-1}$\,Mpc. The correlation weakens at larger scales ($> 6\,h^{-1}$\,Mpc) and becomes noisier at smaller scales ($< 2\,h^{-1}$\,Mpc), indicating that large-scale structure primarily drive environmental effects on AGN activity.
	
\end{enumerate}

Our results have important implications for understanding AGN triggering mechanisms and galaxy evolution in different environments. The strong environmental dependence of X-ray AGN, combined with the lack of such dependence in WISE and BL AGN selections, suggests that obscured AGN activity (more readily detected in X-rays) is preferentially triggered in high-density environments, possibly through mechanisms such as galaxy harassment, tidal interactions, or an enhanced merger rate. The absence of Eddington ratio variations indicates that the available gas supply and feeding mechanisms are similar across environments once an AGN is triggered, pointing to environmental effects primarily operating on the frequency of triggering events rather than on the subsequent accretion physics.

The $\delta_{\rm rank}$ methodology introduced here provides a powerful tool for future studies to cleanly separate stellar mass and environmental effects in large galaxy surveys. The strong SFR-environment-AGN connection we uncover, especially the extreme X-ray AGN fractions ($\sim\!60\%$) in intermediate-density, high-SFR galaxies, warrants further investigation with higher spectral resolution observations and detailed morphological studies to identify the physical processes responsible for this enhanced activity.

Future work combining these statistical results with resolved observations of individual galaxies in different environments will be crucial for understanding the detailed physical mechanisms by which environment modulates AGN triggering while leaving the fundamental accretion process unchanged.

\section{Data Availability}
\label{sec:data_avail}
Our pipeline for performing the analysis is currently publicly available at \url{gitlab-link-to-gamaerositaproject}. Most value added data will be made publicly available on above repository upon publication.

\section*{Acknowledgments}

 This work was supported by the Department of Atomic Energy, Government of India, under Project Identification Number RTI-4012. The computations were carried out on the computing clusters at the Department of Theoretical Physics, TIFR, Mumbai. We thank Kapil Ghadiali and Ajay Salve for their support with the HPC facility at TIFR, Mumbai. 
 
NS acknowledges TIFR-DTP for financial support during multiple long-term visits to TIFR and the Ministry of Education (MoE), Government of India, for fellowship support since February 2026. NS also acknowledges useful discussions with Suchira Sarkar and thanks the other members of the Physics of Universe and Galaxies (PUG) group at TIFR, Mumbai, for many stimulating discussions. It is also a pleasure to acknowledge many valuable, free-flowing discussions with the participants of the Pune-Mumbai Cosmology and Astroparticle Physics (PM-CAP) series of meetings.\footnote{\url{https://www.tifr.res.in/~shadab.alam/PM_CAP_meeting/}}

GAMA is a joint European-Australasian project based around a spectroscopic
campaign using the Anglo-Australian Telescope. The GAMA input catalogue is
based on data taken from the Sloan Digital Sky Survey and the UKIRT Infrared
Deep Sky Survey. Complementary imaging of the GAMA regions is being obtained
by a number of independent survey programmes including GALEX MIS, VST KiDS,
VISTA VIKING, WISE, Herschel-ATLAS, GMRT and ASKAP providing UV to radio
coverage. GAMA is funded by the STFC (UK), the ARC (Australia), the AAO,
and the participating institutions. The GAMA website is
\url{http://www.gama-survey.org/}.

This work is based on data from eROSITA, the soft X-ray instrument aboard
SRG, a joint Russian-German science mission supported by the Russian Space
Agency (Roskosmos), in the interests of the Russian Academy of Sciences
represented by its Space Research Institute (IKI), and the Deutsches
Zentrum f\"ur Luft- und Raumfahrt (DLR). The SRG spacecraft was built by
Lavochkin Association (NPOL) and its subcontractors, and is operated by
NPOL with support from the Max Planck Institute for Extraterrestrial
Physics (MPE). The development and construction of the eROSITA X-ray
instrument was led by MPE, with contributions from the Dr. Karl Remeis
Observatory Bamberg \& ECAP (FAU Erlangen-Nuernberg), the University of
Hamburg Observatory, the Leibniz Institute for Astrophysics Potsdam (AIP),
and the Institute for Astronomy and Astrophysics of the University of
T\"ubingen, with the support of DLR and the Max Planck Society. The
Argelander Institute for Astronomy of the University of Bonn and the
Ludwig Maximilians Universit\"at Munich also participated in the science
preparation for eROSITA. 

This publication makes use of data products from the Wide-field Infrared Survey Explorer, 
which is a joint project of the University of California, Los Angeles, and the Jet Propulsion 
Laboratory/California Institute of Technology, and NEOWISE, which is a project of the Jet 
Propulsion Laboratory/California Institute of Technology. WISE and NEOWISE are funded by 
the National Aeronautics and Space Administration.

The DESI Legacy Imaging Surveys consist of three individual and
complementary projects: the Dark Energy Camera Legacy Survey (DECaLS),
the Beijing-Arizona Sky Survey (BASS), and the Mayall z-band Legacy
Survey (MzLS). DECaLS, BASS and MzLS together include data obtained,
respectively, at the Blanco telescope, Cerro Tololo Inter-American
Observatory, NSF's NOIRLab; the Bok telescope, Steward Observatory,
University of Arizona; and the Mayall telescope, Kitt Peak National
Observatory, NOIRLab. NOIRLab is operated by the Association of
Universities for Research in Astronomy (AURA) under a cooperative
agreement with the National Science Foundation. Pipeline processing and
analyses of the data were supported by NOIRLab and the Lawrence Berkeley
National Laboratory (LBNL). Legacy Surveys also uses data products from
the Near-Earth Object Wide-field Infrared Survey Explorer (NEOWISE), a
project of the Jet Propulsion Laboratory/California Institute of
Technology, funded by the National Aeronautics and Space Administration.
Legacy Surveys was supported by: the Director, Office of Science, Office
of High Energy Physics of the U.S. Department of Energy; the National
Energy Research Scientific Computing Center, a DOE Office of Science User
Facility; the U.S. National Science Foundation, Division of Astronomical
Sciences; the National Astronomical Observatories of China, the Chinese
Academy of Sciences and the Chinese National Natural Science Foundation.
LBNL is managed by the Regents of the University of California under
contract to the U.S. Department of Energy. The complete acknowledgments
can be found at \url{https://www.legacysurvey.org/acknowledgment/}.

This research used data obtained with the Dark Energy Spectroscopic
Instrument (DESI). DESI construction and operations is managed by the
Lawrence Berkeley National Laboratory. This material is based upon work
supported by the U.S. Department of Energy, Office of Science, Office of
High-Energy Physics, under Contract No. DE--AC02--05CH11231, and by the
National Energy Research Scientific Computing Center, a DOE Office of
Science User Facility under the same contract. Additional support for
DESI was provided by the U.S. National Science Foundation, Division of
Astronomical Sciences under Contract No. AST-0950945 to the NSF's
National Optical-Infrared Astronomy Research Laboratory; the Science and
Technology Facilities Council of the United Kingdom; the Gordon and Betty
Moore Foundation; the Heising-Simons Foundation; the French Alternative
Energies and Atomic Energy Commission (CEA); the National Council of
Humanities, Science and Technology of Mexico (CONAHCYT); the Ministry of
Science, Innovation and Universities of Spain (MICIU/AEI/10.13039/501100011033),
and by the DESI Member Institutions:
\url{https://www.desi.lbl.gov/collaborating-institutions}. The DESI
collaboration is honored to be permitted to conduct scientific research
on I'oligam Du'ag (Kitt Peak), a mountain with particular significance to
the Tohono O'odham Nation. Any opinions, findings, and conclusions or
recommendations expressed in this material are those of the author(s) and
do not necessarily reflect the views of the U.S. National Science
Foundation, the U.S. Department of Energy, or any of the listed funding
agencies.

This work made extensive use of the open-source computing packages NumPy \citep{vanderwalt-numpy},\footnote{\url{http://www.numpy.org}} SciPy \citep{scipy},\footnote{\url{http://www.scipy.org}} Pandas \citep{mckinney-proc-scipy-2010, reback2020pandas},\footnote{\url{https://pandas.pydata.org}} Matplotlib \citep{hunter07_matplotlib},\footnote{\url{https://matplotlib.org/}} and Jupyter Notebook.\footnote{\url{https://jupyter.org}} 
The analysis was performed on the Pawna cluster at TIFR, Mumbai. Machine learning tools were used to suggest language improvements and gather references within the manuscript.

\facilities{GAMA, eROSITA:eFEDS, WISE, GALEX, Legacy Survey, DESI}

\bibliographystyle{aasjournal}
\bibliography{references}

\end{document}